\documentclass[12pt]{report}
\usepackage[all]{xy}

\advance \topmargin by -\headheight
\advance \topmargin by -\headsep
     
\evensidemargin \oddsidemargin
\begin{document}

\topmargin -.6in
\def\nonu{\nonumber}
\def\rf#1{(\ref{eq:#1})}
\def\lab#1{\label{eq:#1}} 
\def\br{\begin{eqnarray}}
\def\er{\end{eqnarray}}
\def\be{\begin{equation}}
\def\ee{\end{equation}}
\def\0{\nonumber}
\def\lb{\lbrack}
\def\rb{\rbrack}
\def\({\left(}
\def\){\right)}
\def\v{\vert}
\def\bv{\bigm\vert}
\def\lskip{\vskip\baselineskip\vskip-\parskip\noindent}
\relax
\newcommand{\nit}{\noindent}
\newcommand{\ct}[1]{\cite{#1}}
\newcommand{\bi}[1]{\bibitem{#1}}
\def\a{\alpha}
\def\b{\beta}
\def\ca{{\cal A}}
\def\cm{{\cal M}}
\def\cn{{\cal N}}
\def\cf{{\cal F}}
\def\d{\delta} 
\def\D{\Delta}
\def\eps{\epsilon}
\def\g{\gamma}
\def\G{\Gamma}
\def\grad{\nabla}
\def\h{ {1\over 2}  }
\def\hc{\hat{c}}
\def\hd{\hat{d}}
\def\hg{\hat{g}}
\def\hp{ {+{1\over 2}}  }
\def\hm{ {-{1\over 2}}  }
\def\k{\kappa}
\def\l{\lambda}
\def\L{\Lambda}
\def\lg{\langle}
\def\m{\mu}
\def\n{\nu}
\def\o{\over}
\def\om{\omega}
\def\O{\Omega}
\def\p{\phi}
\def\pa{\partial}
\def\pr{\prime}
\def\ra{\rightarrow}
\def\rh{\rho}
\def\rg{\rangle}
\def\s{\sigma}
\def\t{\tau}
\def\th{\theta}
\def\ti{\tilde}
\def\wti{\widetilde}
\def\inte{\int dx }
\def\xb{\bar{x}}
\def\yb{\bar{y}}

\def\tr{\mathop{\rm tr}}
\def\Tr{\mathop{\rm Tr}}
\def\partder#1#2{{\partial #1\over\partial #2}}
\def\ds{{\cal D}_s}
\def\wtwo{{\wti W}_2}
\def\lie{{\cal G}}
\def\alie{{\widehat \lie}}
\def\dlie{{\cal G}^{\ast}}
\def\elie{{\widetilde \lie}}
\def\edlie{{\elie}^{\ast}}
\def\hlie{{\cal H}}
\def\wlie{{\widetilde \lie}}

\def\rlx{\relax\leavevmode}
\def\inbar{\vrule height1.5ex width.4pt depth0pt}
\def\IZ{\rlx\hbox{\sf Z\kern-.4em Z}}
\def\IR{\rlx\hbox{\rm I\kern-.18em R}}
\def\IC{\rlx\hbox{\,$\inbar\kern-.3em{\rm C}$}}
\def\one{\hbox{{1}\kern-.25em\hbox{l}}}

\def\PRL#1#2#3{{\sl Phys. Rev. Lett.} {\bf#1} (#2) #3}
\def\NPB#1#2#3{{\sl Nucl. Phys.} {\bf B#1} (#2) #3}
\def\NPBFS#1#2#3#4{{\sl Nucl. Phys.} {\bf B#2} [FS#1] (#3) #4}
\def\CMP#1#2#3{{\sl Commun. Math. Phys.} {\bf #1} (#2) #3}
\def\PRD#1#2#3{{\sl Phys. Rev.} {\bf D#1} (#2) #3}
\def\PLA#1#2#3{{\sl Phys. Lett.} {\bf #1A} (#2) #3}
\def\PLB#1#2#3{{\sl Phys. Lett.} {\bf #1B} (#2) #3}
\def\JMP#1#2#3{{\sl J. Math. Phys.} {\bf #1} (#2) #3}
\def\PTP#1#2#3{{\sl Prog. Theor. Phys.} {\bf #1} (#2) #3}
\def\SPTP#1#2#3{{\sl Suppl. Prog. Theor. Phys.} {\bf #1} (#2) #3}
\def\AoP#1#2#3{{\sl Ann. of Phys.} {\bf #1} (#2) #3}
\def\PNAS#1#2#3{{\sl Proc. Natl. Acad. Sci. USA} {\bf #1} (#2) #3}
\def\RMP#1#2#3{{\sl Rev. Mod. Phys.} {\bf #1} (#2) #3}
\def\PR#1#2#3{{\sl Phys. Reports} {\bf #1} (#2) #3}
\def\AoM#1#2#3{{\sl Ann. of Math.} {\bf #1} (#2) #3}
\def\UMN#1#2#3{{\sl Usp. Mat. Nauk} {\bf #1} (#2) #3}
\def\FAP#1#2#3{{\sl Funkt. Anal. Prilozheniya} {\bf #1} (#2) #3}
\def\FAaIA#1#2#3{{\sl Functional Analysis and Its Application} {\bf #1} (#2)
#3}
\def\BAMS#1#2#3{{\sl Bull. Am. Math. Soc.} {\bf #1} (#2) #3}
\def\TAMS#1#2#3{{\sl Trans. Am. Math. Soc.} {\bf #1} (#2) #3}
\def\InvM#1#2#3{{\sl Invent. Math.} {\bf #1} (#2) #3}
\def\LMP#1#2#3{{\sl Letters in Math. Phys.} {\bf #1} (#2) #3}
\def\IJMPA#1#2#3{{\sl Int. J. Mod. Phys.} {\bf A#1} (#2) #3}
\def\AdM#1#2#3{{\sl Advances in Math.} {\bf #1} (#2) #3}
\def\RMaP#1#2#3{{\sl Reports on Math. Phys.} {\bf #1} (#2) #3}
\def\IJM#1#2#3{{\sl Ill. J. Math.} {\bf #1} (#2) #3}
\def\APP#1#2#3{{\sl Acta Phys. Polon.} {\bf #1} (#2) #3}
\def\TMP#1#2#3{{\sl Theor. Mat. Phys.} {\bf #1} (#2) #3}
\def\JPA#1#2#3{{\sl J. Physics} {\bf A#1} (#2) #3}
\def\JSM#1#2#3{{\sl J. Soviet Math.} {\bf #1} (#2) #3}
\def\MPLA#1#2#3{{\sl Mod. Phys. Lett.} {\bf A#1} (#2) #3}
\def\JETP#1#2#3{{\sl Sov. Phys. JETP} {\bf #1} (#2) #3}
\def\JETPL#1#2#3{{\sl  Sov. Phys. JETP Lett.} {\bf #1} (#2) #3}
\def\PHSA#1#2#3{{\sl Physica} {\bf A#1} (#2) #3}
\def\PHSD#1#2#3{{\sl Physica} {\bf D#1} (#2) #3}


\begin{titlepage}

\vskip 1 cm
\begin{center}
{\Large\bf Permutability of Backlund Transformations for $N=2$ Supersymmetric Sine-Gordon  } \vglue 1  true cm

\vspace{.8 cm}
J.F. Gomes,  L.H. Ymai and A.H. Zimerman \\
\vspace{.35 cm}
{\footnotesize Instituto de Física Teórica,
IFT-UNESP\\
Universidade Estadual Paulista\\
Caixa Postal 70532-2\\
01156-970 São Paulo, SP, Brazil\\}

\vspace{1 cm}

\end{center}

\normalsize
\vskip 0.2cm

\begin{center}
{\large {\bf ABSTRACT}}\\
\end{center}
\noindent

The permutability of two Backlund transformations  is employed  
to construct a non linear superposition formula  and to generate  a class of solutions for the $N=2$ super sine-Gordon model.  We present explicitly the one and two soliton solutions.

\noindent

\vglue 1 true cm

\end{titlepage}

\section*{1\hspace{0.5cm}Introduction}
 Backlund transformations reduce the order of the non-linear differential equations making the system  sometimes effectively more tractable.  Starting with a simple input solution, we may be able to solve for a more complicated one.  In many cases, this may be very difficult to accomplish.  A convenient and powerful  way is to  use the {\it permutability theorem } which provides a  closed algebraic  non-linear superposition formula for the solutions.

 The Backlund transformation and the Permutability theorem  are employed to derive a series of consistency conditions which are satisfied by soliton solutions of certain class of integrable models.  Within such class, we encounter the   sine-Gordon  \cite{rogers} and KdV  \cite{wahlkist} equations.  This framework  was also applied to the $N=1$  super KdV \cite{liu} and super sinh-Gordon \cite{n1}  in order to derive its  soliton solutions.
 
 The $N=2$ super sine-Gordon model was proposed in \cite{uematsu} and  later in \cite{unpublished} its algebraic structure was uncovered.
  Certain solutions  of this model have already been constructed \cite{n2}, however they were such that involve a single  Grassmann parameter.  
 In this paper we extend the non-linear superposition formulae  for soliton solutions  of the $N=2$ super sine-Gordon model.  These formulae are derived  from the Backlund transformation proposed in \cite{ymai1} and the permutability condition which implies that the order of  two successive Backlund transformations  is irrelevant.
 As examples, we present explicitly the 1-and 2-soliton solutions with distinct Grassmann parameters.
 
 Recently  the Pohlmeyer reduction of $AdS_n x S_n$ superstring models have been considered \cite{1}  which in the simple case of $n=2$ was shown  \cite{2} to be equivalent to the $N=2$ supersymmetric sine-Gordon.

 This paper is organized as follows.  In Section 2 we discuss the $N=2$ super sine-Gordon  and its  Backlund Transformation.  In Section 3 we apply the permutability  condition to derive a  closed algebraic  non-linear superposition formulae  involving  solutions of the model.  Finally in Section 4 and 5 we present the 1- and 2-soliton solutions respectively.  In the appendix A we present the Backlund transformation in components.  In appendices B and C we give details for the derivation of the superposition formulae.

\section*{2\hspace{0.5cm}$N=2$ super sine-Gordon - Backlund Transformation}

Let us start by introducing the $N=2$ superfields \cite{uematsu}
\begin{eqnarray}
\phi^{\pm}&=&\varphi^{\pm}(z^{\pm},\bar{z}^{\pm})+\theta^{\pm}\psi^{\mp}(z^{\pm},\bar{z}^{\pm})+\bar{\theta}^{\pm}\bar{\psi}^{\mp}(z^{\pm},\bar{z}^{\pm})+\theta^{\pm}\bar{\theta}^{\pm}F^{\pm}(z^{\pm},\bar{z}^{\pm}),\nonumber
\end{eqnarray}
where
\begin{eqnarray}
z^{\pm}=z\pm\frac{1}{2}\theta^{+}\theta^{-}, \qquad \bar{z}^{\pm}&=&\bar{z}\pm\frac{1}{2}\bar{\theta}^{+}\bar{\theta}^{-}.\nonumber
\end{eqnarray}

The superfield components 
$\phi^{\pm}$ can be expanded in Grassmann variables  $\theta^{\pm}$ and $\bar{\theta}^{\pm}$.  For instance, the  component $\varphi^{\pm}(z^{\pm},\bar{z}^{\pm})$  gives rise to 
\begin{eqnarray}
\varphi^{\pm}(z^\pm,\bar{z}^\pm)&=&\varphi^\pm \pm\frac{1}{2}\theta^+\theta^-\partial_z\varphi^\pm \pm \frac{1}{2}\bar{\theta}^+\bar{\theta}^-\partial_{\bar{z}}\varphi^\pm +\frac{1}{4}\theta^+\theta^-\bar{\theta}^+\bar{\theta}^-\partial_z\partial_{\bar{z}}\varphi^\pm.\nonumber
\end{eqnarray}
By expanding all components of 
 $\phi^{\pm}$,we obtain
\begin{eqnarray}
\phi^\pm&=&\varphi^\pm+\theta^\pm\psi^\mp +\bar{\theta}^\pm\bar{\psi}^\mp\pm\frac{1}{2}\theta^+\theta^-\partial_z\varphi^\pm \pm\frac{1}{2}\bar{\theta}^+\bar{\theta}^-\partial_{\bar{z}}\varphi^\pm +\theta^\pm\bar{\theta}^\pm F^\pm \nonumber\\
&&\pm\theta^\pm\bar{\theta}^+\bar{\theta}^-\frac{1}{2}\partial_{\bar{z}}\psi^\mp \pm \bar{\theta}^\pm\theta^+\theta^-\frac{1}{2}\partial_z\bar{\psi}^\mp+\frac{1}{4}\theta^+\theta^-\bar{\theta}^+\bar{\theta}^-\partial_z\partial_{\bar{z}}\varphi^\pm.\nonumber
\end{eqnarray}
We next introduce the super derivatives
\begin{eqnarray}
D_{\pm}=\frac{\partial}{\partial\theta^{\pm}}+\frac{1}{2}\theta^{\mp}\partial_{z},
\qquad
\bar{D}_{\pm}=\frac{\partial}{\partial\bar{\theta}^{\pm}}+\frac{1}{2}\bar{\theta}^{\mp}\partial_{\bar{z}},\nonumber
\end{eqnarray}
satisfying the following conditions
\begin{eqnarray}
&&D_{\pm}^2=0, \qquad \qquad \quad \,\, \bar{D}_{\pm}^2=0,\nonumber\\
&&\{\bar{D}_{\pm},D_{\pm}\}=0,\qquad \,\, \{\bar{D}_{\pm},D_{\mp}\}=0,\nonumber\\
&&\{D_{+},D_{-}\}=\partial_z, \qquad \{\bar{D}_{+},\bar{D}_{-}\}=\partial_{\bar{z}}.\nonumber
\end{eqnarray}
The equations of motion for the supersymmetric sine-Gordon model with $N=2$ are given by \cite{uematsu}
\begin{eqnarray}
\bar{D}_{\pm}D_{\pm}\phi^{\pm}=g\, \sin\left(\beta\phi^{\mp}\right),\label{SSGN2}
\end{eqnarray}
where $g$ is a mass parameter and  $\beta$ is the coupling constant. From now on we assume $\b =1$ 
which may be re-inserted by a convenient field reparametrization.  In components, the equations of motion  for the $N=2$ super sine-Gordon reads,
\begin{eqnarray}
F^{\pm}&=&g\, \sin\varphi^{\mp},\nonumber\\
\partial_{\bar{z}}\psi^{\mp}&=&g\,\cos\varphi^{\mp}\bar{\psi}^{\pm},\nonumber\\
\partial_{z}\bar{\psi}^{\mp}&=&-g\,\cos\varphi^{\mp}\psi^{\pm},\nonumber\\
\partial_{z}\partial_{\bar{z}}\varphi^{\pm}&=&-g\,\cos\varphi^{\mp}F^{\mp}-g\,\sin\varphi^{\mp}\psi^{\pm}\bar{\psi}^{\pm}.\nonumber
\end{eqnarray}
Moreover the the chiral, $\phi^{+}$ and the anti-chiral, $\phi^{-}$ superfields satisfy the conditions
\begin{eqnarray}\label{quiral}
\bar{D}_{\pm}\phi^{\mp}=D_{\pm}\phi^{\mp}=0. 
\end{eqnarray}

Let us now recall the Backlund transformation for the $N=2$ super sine-Gordon model \cite{ymai1}.  For this purpose, consider the pair of first order differential equations 
\begin{eqnarray}
D_{+}\phi_{1}^{+}&=&D_{+}\phi_{2}^{+}-\frac{8}{\kappa}\mathcal{F}\cos\left(\frac{\phi_{1}^{-}+\phi_{2}^{-}}{2}\right),\label{f1}\\
\bar{D}_{+}\phi_{1}^{+}&=&-\bar{D}_{+}\phi_{2}^{+}+\kappa\mathcal{G}\cos\left(\frac{\phi_{1}^{-}-\phi_{2}^{-}}{2}\right),\label{f2}
\end{eqnarray}
where $\mathcal{F}$ and $\mathcal{G}$ are fermionic auxiliary superfields
and $\kappa$ is an arbitrary constant.  The above equation and the condition
\begin{eqnarray}
(\bar{D}_{+}D_{+}+D_{+}\bar{D}_{+})\phi_{1}^{+}=0,\nonumber
\end{eqnarray}
leads to the equations of motion 
\begin{eqnarray}
\bar{D}_{+}D_{+}\phi_{2}^{+}=g\, \sin\phi_{2}^{-},\nonumber
\end{eqnarray}
provided the superfields 
 $\mathcal{F}$ and  $\mathcal{G}$ satisfy
\begin{eqnarray}
\bar{D}_{+}\mathcal{F}=-\frac{\kappa g}{4}\sin\left(\frac{\phi_{1}^{-}-\phi_{2}^{-}}{2}\right),\qquad
D_{+}\mathcal{G}=-\frac{2g}{\kappa}\sin\left(\frac{\phi_{1}^{-}+\phi_{2}^{-}}{2}\right).\label{F1F2}
\end{eqnarray}
In a similar way,  
\begin{eqnarray}
D_{-}\phi_{1}^{-}&=&D_{-}\phi_{2}^{-}+\lambda\mathcal{G}\cos\left(\frac{\phi_{1}^{+}+\phi_{2}^{+}}{2}\right),\label{g1}\\
\bar{D}_{-}\phi_{1}^{-}&=&-\bar{D}_{-}\phi_{2}^{-}-\frac{8}{\lambda}\mathcal{F}\cos\left(\frac{\phi_{1}^{+}-\phi_{2}^{+}}{2}\right),\label{g2}
\end{eqnarray}
where $\lambda$ is another arbitrary constant.   Together with the condition 
\begin{eqnarray}
(\bar{D}_{-}D_{-}+D_{-}\bar{D}_{-})\phi_{1}^{-}=0,\nonumber
\end{eqnarray}
yields
\begin{eqnarray}
\bar{D}_{-}D_{-}\phi_{2}^{-}=g\, \sin\phi_{2}^{+},\nonumber
\end{eqnarray}
provided  $\mathcal{G}$ and  $\mathcal{F}$ satisfy 
\begin{eqnarray}
\bar{D}_{-}\mathcal{G}=\frac{2g}{\lambda}\sin\left(\frac{\phi_{1}^{+}-\phi_{2}^{+}}{2}\right), \qquad
D_{-}\mathcal{F}=\frac{\lambda g}{4}\sin\left(\frac{\phi_{1}^{+}+\phi_{2}^{+}}{2}\right).\label{G1G2}
\end{eqnarray}
Acting with $D_+$ in  eqn. (\ref{f1}), $\bar{D}_+$ in (\ref{f2}), $D_-$ in (\ref{g1}) and  $\bar{D}_-$ in (\ref{g2}) we find
\begin{eqnarray}\label{conditions}
D_{+}{\mathcal{F}}=0, \qquad \bar{D}_+{\mathcal{G}}=0, \qquad D_-{\mathcal{G}}=0, \qquad \bar{D}_{-}{\mathcal{F}}=0.   
\end{eqnarray}
These last conditions allows us to rewrite the fermionic superfields into two distinct manners, i.e.,
\begin{eqnarray}\label{DPhi}
\mathcal{F}&=&D_{+}\Phi_{1}^{+}=\bar{D}_{-}\Phi_{2}^{-},\label{F}\\
\mathcal{G}&=&D_{-}\Phi_{1}^{-}=\bar{D}_{+}\Phi_{2}^{+},\label{G} 
\end{eqnarray}
where the chiral  $\Phi_p^{+}$  and anti-chiral $\Phi_p^{-}, p=1,2$ superfields are defined as
\begin{eqnarray}
\Phi_{1}^{\pm}&=&q_{1}^{\pm}(z^{\pm},\bar{z}^{\pm})+\theta^{\pm}\zeta_{1}^{\pm}(z^{\pm},\bar{z}^{\pm})+\bar{\theta}^{\pm}\zeta_{2}^{\pm}(z^{\pm},\bar{z}^{\pm})+\theta^{\pm}\bar{\theta}^{\pm}q_{2}^{\pm}(z^{\pm},\bar{z}^{\pm}),\nonumber\\
\Phi_{2}^{\pm}&=&p_{1}^{\pm}(z^{\pm},\bar{z}^{\pm})+\theta^{\pm}\xi_{1}^{\pm}(z^{\pm},\bar{z}^{\pm})+\bar{\theta}^{\pm}\xi_{2}^{\pm}(z^{\pm},\bar{z}^{\pm})+\theta^{\pm}\bar{\theta}^{\pm}p_{2}^{\pm}(z^{\pm},\bar{z}^{\pm}).\nonumber
\end{eqnarray}
The second equality in  (\ref{F}) implies
\begin{eqnarray}
\zeta_{1}^{+}=\xi_{2}^{-}, \qquad
q_{2}^{+}=\partial_{\bar{z}}p_{1}^{-}, \qquad
p_{2}^{-}=-\partial_{z}q_{1}^{+}, \qquad
\partial_{z}\zeta_{2}^{+}=-\partial_{\bar{z}}\xi_{1}^{-},\label{f1g2}
\end{eqnarray} 
whilst  the second equality in 
(\ref{G}) implies
\begin{eqnarray}
\zeta_{1}^{-}=\xi_{2}^{+}, \qquad p_{2}^{+}=-\partial_{z}q_{1}^{-},
\qquad q_{2}^{-}=\partial_{\bar{z}}p_{1}^{+}, \qquad
\partial_{z}\zeta_{2}^{-}=-\partial_{\bar{z}}\xi_{1}^{+}.\label{f2g1}
\end{eqnarray}

Eqns. (\ref{f1})-(\ref{conditions}) describe the Backlund transformation for the $N=2$ super sine-Gordon system.  In appendix A we present these equations in components.

\section*{3\hspace{0.5cm}The Permutability condition}

A Backlund transformation from $\phi_0^{\pm}$ to $\phi_1^{\pm}$ is described by 
\begin{eqnarray}
D_+(\phi_0^+-\phi_1^+)&=&-\frac{8}{\kappa_1}\mathcal{F}^{(0,1)}\cos\left(\frac{\phi_0^-+\phi_1^-}{2}\right),\label{eq1}\\
\bar{D}_+(\phi_0^++\phi_1^+)&=&\kappa_1 \mathcal{G}^{(0,1)}\cos\left(\frac{\phi_0^--\phi_1^-}{2}\right),\label{eq2}\\
D_-(\phi_0^--\phi_1^-)&=&\lambda_1\mathcal{G}^{(0,1)}\cos\left(\frac{\phi_0^++\phi_1^+}{2}\right),\label{eq3}\\
\bar{D}_-(\phi_0^-+\phi_1^-)&=&-\frac{8}{\lambda_1} \mathcal{F}^{(0,1)}\cos\left(\frac{\phi_0^+-\phi_1^+}{2}\right),\label{eq4}
\end{eqnarray}
where we have introduced the superscript indices $(0,1)$   for the auxiliary fermionic superfields denoting its dependence in $\phi_0^{\pm}$ and  $\phi_1^{\pm}$.  The later, in turn satisfy the following condition (as in (\ref{F1F2}) and (\ref{G1G2}))
\begin{eqnarray}
\bar{D}_+\mathcal{F}^{(0,1)}&=&-g\frac{\kappa_1}{4}\sin\left(\frac{\phi_0^--\phi_1^-}{2}\right),\label{p1}\\
D_+\mathcal{G}^{(0,1)}&=&-g\frac{2}{\kappa_1}\sin\left(\frac{\phi_0^-+\phi_1^-}{2}\right),\label{p2}\\
\bar{D}_-\mathcal{G}^{(0,1)}&=&g\frac{2}{\lambda_1}\sin\left(\frac{\phi_0^+-\phi_1^+}{2}\right),\label{p3}\\
D_-\mathcal{F}^{(0,1)}&=&g\frac{\lambda_1}{4}\sin\left(\frac{\phi_0^++\phi_1^+}{2}\right).\label{p4}
\end{eqnarray}
The chiral conditions (\ref{conditions}), i.e.,
\begin{eqnarray}
\bar{D}_-{\mathcal{F}}^{(0,1)}=0, \qquad D_+{\mathcal{F}}^{(0,1)}=0, \qquad \bar{D}_+{\mathcal{G}}^{(0,1)}=0, \qquad D_-{\mathcal{G}}^{(0,1)}=0,\nonumber
\end{eqnarray} 
are automatically satisfied  using eqn. (\ref{DPhi}), i.e., expressing the fermionic superfields as derivatives of chiral superfields, 
\begin{eqnarray}
\mathcal{F}^{(0,1)}=D_+\Phi_1^{+(0,1)}=\bar{D}_-\Phi_2^{-(0,1)},\qquad
\mathcal{G}^{(0,1)}=D_-\Phi_1^{-(0,1)}=\bar{D}_+\Phi_2^{+(0,1)},\nonumber 
\end{eqnarray}
where the superscript indices  indicate whether the superfield 
 $\Phi_1^{\pm}$ and  $\Phi_2^{\pm}$ depend upon $\phi_0^{\pm}$ and $\phi_1^{\pm}$.
Acting with super derivatives  $D_-, \bar D_-, D_+$ and $\bar D_+$ on eqns.(\ref{eq1}), (\ref{eq2}), (\ref{eq3}) and  (\ref{eq4}) respectively, we find 
\begin{eqnarray}
\partial_z(\phi_0^+-\phi_1^+)&=&-2\gamma_1\, s_{0,1}^+c_{0,1}^-+\frac{8}{\kappa_1}{\mathcal{F}}^{(0,1)}D_-c_{0,1}^-,\label{equation1}\\
\partial_{\bar{z}}(\phi_0^++\phi_1^+)&=&2\frac{g^2}{\gamma_1}\,\bar{s}_{0,1}^+\bar{c}_{0,1}^--\kappa_1{\mathcal{G}}^{(0,1)}\bar{D}_-\bar{c}_{0,1}^-,\label{equation2}\\
\partial_{z}(\phi_0^--\phi_1^-)&=&-2\gamma_1\, s_{0,1}^- c_{0,1}^+-\lambda_1{\mathcal{G}}^{(0,1)}D_+ c_{0,1}^+,\label{equation3}\\
\partial_{\bar{z}}(\phi_0^-+\phi_1^-)&=&2\frac{g^2}{\gamma_1} \bar{s}_{0,1}^-\bar{c}_{0,1}^++\frac{8}{\lambda_1}{\mathcal{F}}^{(0,1)}\bar{D}_+\bar{c}_{0,1}^+,\label{equation4}
\end{eqnarray}
where $\gamma_1=g\frac{\lambda_1}{\kappa_1}$ and 
\begin{eqnarray}
c_{j,k}^{\pm}&=&\cos\left(\frac{\phi_j^{\pm}+\phi_k^{\pm}}{2}\right), \qquad s_{j,k}^{\pm}=\sin\left(\frac{\phi_j^{\pm}+\phi_k^{\pm}}{2}\right),\label{x1}
\\
\bar{c}_{j,k}^{\pm}&=&\cos\left(\frac{\phi_j^{\pm}-\phi_k^{\pm}}{2}\right), \qquad \bar{s}_{j,k}^{\pm}=\sin\left(\frac{\phi_j^{\pm}-\phi_k^{\pm}}{2}\right),\label{x2}
\end{eqnarray}

We now assume that the order of two successive Backlund transformations  is irrelevant leading to the same final result.  
Such condition is known as the permutability theorem, i.e.,   
$\xymatrix@1{\phi_0^{\pm}\ar[r]^(.5){\gamma_1}&\phi_1^{\pm}\ar[r]^(.5){\gamma_2}&\phi_{12}^{\pm}}$ and in the inverse order, $\xymatrix@1{\phi_0^{\pm}\ar[r]^(.5){\gamma_2}&\phi_2^{\pm}\ar[r]^(.5){\gamma_1}&\phi_{21}^{\pm}}$, does not change the final result, $\phi_{12}^{\pm}=\phi_{21}^{\pm}\equiv \phi_3^{\pm}$.

The permutability theorem applied to the  Backlund equation (\ref{eq1}) leads to
\begin{eqnarray}
D_+(\phi_0^+-\phi_1^+)&=&-\frac{8}{\kappa_1}\mathcal{F}^{(0,1)}c_{0,1}^-,\nonumber\\
D_+(\phi_1^+-\phi_3^+)&=&-\frac{8}{\kappa_2}\mathcal{F}^{(1,3)}c_{1,3}^-,\nonumber\\
D_+(\phi_0^+-\phi_2^+)&=&-\frac{8}{\kappa_2}\mathcal{F}^{(0,2)}c_{0,2}^-,\nonumber\\
D_+(\phi_2^+-\phi_3^+)&=&-\frac{8}{\kappa_1}\mathcal{F}^{(2,3)}c_{2,3}^-.\label{D+}
\end{eqnarray} 
Taking into account that  the sum of the  first two and the last two  equations are the same, we obtain,
\begin{eqnarray}
\frac{1}{\kappa_1}\mathcal{F}^{(0,1)}c_{0,1}^-+\frac{1}{\kappa_2}\mathcal{F}^{(1,3)}c_{1,3}^-=\frac{1}{\kappa_2}\mathcal{F}^{(0,2)}c_{0,2}^-+\frac{1}{\kappa_1}\mathcal{F}^{(2,3)}c_{2,3}^-.\label{relation1}
\end{eqnarray}
Similarly, from (\ref{eq4}), we obtain
\begin{eqnarray}
\bar{D}_-(\phi_0^-+\phi_1^-)&=&-\frac{8}{\lambda_1} \mathcal{F}^{(0,1)}\bar{c}_{0,1}^+,\nonumber\\
\bar{D}_-(\phi_1^-+\phi_3^-)&=&-\frac{8}{\lambda_2} \mathcal{F}^{(1,3)}\bar{c}_{1,3}^+,\nonumber\\
\bar{D}_-(\phi_0^-+\phi_2^-)&=&-\frac{8}{\lambda_2} \mathcal{F}^{(0,2)}\bar{c}_{0,2}^+,\nonumber\\
\bar{D}_-(\phi_2^-+\phi_3^-)&=&-\frac{8}{\lambda_1} \mathcal{F}^{(2,3)}\bar{c}_{2,3}^+,\label{D-}
\label{xx}
\end{eqnarray}
leading to 
\begin{eqnarray}
\frac{1}{\lambda_1}\mathcal{F}^{(0,1)}\bar{c}_{0,1}^+-\frac{1}{\lambda_2}\mathcal{F}^{(1,3)}\bar{c}_{1,3}^+=\frac{1}{\lambda_2}\mathcal{F}^{(0,2)}\bar{c}_{0,2}^+-\frac{1}{\lambda_1}\mathcal{F}^{(2,3)}\bar{c}_{2,3}^+.\label{relation2}
\end{eqnarray}
We propose as solution for the non-linear superposition formula 
$\phi_{12}^{\pm}=\phi_{21}^{\pm}=\phi_3^{\pm}$, 
\begin{eqnarray}\label{ansatz}
\phi_3^{\pm}=\phi_0^{\pm}+\Gamma_{\pm}+\Delta_{\pm},   
\end{eqnarray}
with
\begin{eqnarray}
&&\Gamma_{\pm}(x,y)=2\arctan\left[\delta\tan\left(\frac{x+y}{4}\right)\right]\pm 2\arctan\left[\delta\tan\left(\frac{x-y}{4}\right)\right],\nonumber\\
&&x=\phi_1^+-\phi_2^+, \qquad y=\phi_1^--\phi_2^-,\label{Gamma}\\
&&\delta=\frac{\gamma_1+\gamma_2}{\gamma_1-\gamma_2}, \qquad \gamma_k=g\frac{\lambda_k}{\kappa_k}.\nonumber
\end{eqnarray}
Notice that the solution $\phi_3^{\pm}$  when the fermionic superfields are neglected  is derived in the appendix B to be 
 $\phi_3^{\pm}=\phi_0^{\pm}+\Gamma_{\pm}$.  The term $\Delta_{\pm}$ comes from the contribution of the fermionic superfields
 and has the following form
 \begin{eqnarray}
&&\Delta_{\pm}=\sum_{j,k=1}^{2}\Lambda_{j,k}^{\pm}f_{j,k}+\Lambda_{0}^{\pm}f_0,\nonumber\\
&&f_{j,k}=\mathcal{F}^{(0,j)}\mathcal{G}^{(0,k)},\qquad f_0=\mathcal{F}^{(0,1)}\mathcal{F}^{(0,2)}\mathcal{G}^{(0,1)}\mathcal{G}^{(0,2)},\nonumber
\end{eqnarray}     
where we have assumed the coefficients  $\Lambda^{\pm}$ to be functionals of  $x=(\phi_1^{+}-\phi_2^+)$ and  $y=(\phi_1^--\phi_2^-)$,i.e.,  
\begin{eqnarray}
\Lambda_{j,k}^{\pm}=\Lambda_{j,k}^{\pm}(x,y), \qquad \Lambda_{0}^{\pm}=\Lambda_{0}^{\pm}(x,y).
\end{eqnarray}
 Observe that there are no terms  like $\Lambda_1^{\pm}\mathcal{F}^{(0,1)}\mathcal{F}^{(0,2)}$ nor $\Lambda_2^\pm\mathcal{G}^{(0,1)}\mathcal{G}^{(0,2)}$ due to chiral equations (\ref{quiral}).   $\L_{\pm} $ are determined in appendix C where, 
\begin{eqnarray}
\Lambda_{1,1}^+&=&\Lambda_{2,2}^+=-\frac{8\mu_-}{g\eta_+\eta_-}\cos\left(\frac{x}{2}\right)\sin\left(\frac{y}{2}\right),\nonumber\\
\Lambda_{1,2}^+&=&\frac{8\mu_-}{g\eta_+\eta_-}\left(\frac{\lambda_2}{\lambda_1}\right)\sin\left(\frac{y}{2}\right),\nonumber\\
\Lambda_{2,1}^+&=&\frac{8\mu_-}{g\eta_+\eta_-}\left(\frac{\lambda_1}{\lambda_2}\right)\sin\left(\frac{y}{2}\right),\nonumber\\
\Lambda_{0}^+&=&-\frac{32\mu_-}{(g\eta_+\eta_-)^2}\sin\left(\frac{x}{2}\right)\left[\cos\left(\frac{y}{2}\right)(a+\cos x-\cos y)-2\mu_+\cos\left(\frac{x}{2}\right)\right],\nonumber
\end{eqnarray}
\begin{eqnarray}
\Lambda_{1,1}^-&=&\Lambda_{2,2}^-=\frac{8\mu_-}{g\eta_+\eta_-}\cos\left(\frac{y}{2}\right)\sin\left(\frac{x}{2}\right),\nonumber\\
\Lambda_{1,2}^-&=&-\frac{8\mu_-}{g\eta_+\eta_-}\left(\frac{\kappa_2}{\kappa_1}\right)\sin\left(\frac{x}{2}\right),\nonumber\\
\Lambda_{2,1}^-&=&-\frac{8\mu_-}{g\eta_+\eta_-}\left(\frac{\kappa_1}{\kappa_2}\right)\sin\left(\frac{x}{2}\right),\nonumber\\
\Lambda_{0}^-&=&-\frac{32\mu_-}{(g\eta_+\eta_-)^2}\sin\left(\frac{y}{2}\right)\left[\cos\left(\frac{x}{2}\right)(a-\cos x+\cos y)-2\mu_+\cos\left(\frac{y}{2}\right)\right],\nonumber
\end{eqnarray}
\begin{eqnarray}
\mu_{\pm}&=&\frac{\gamma_1}{\gamma_2}\pm\frac{\gamma_2}{\gamma_1}, \nonumber\\ a&=&\frac{1}{2}\left(\frac{\gamma_1^2}{\gamma_2^2}+\frac{\gamma_2^2}{\gamma_1^2}\right)+3, \nonumber\\ \eta_{\pm}&=&\mu_+-2\cos\left(\frac{x\pm y}{2}\right).\nonumber
\end{eqnarray}

\section*{3.1\hspace{0.5cm}Solution in Components}
In components the non-linear superposition formula (\ref{ansatz}) yields the following expressions,
\begin{eqnarray}
\varphi_3^+&=&\varphi_0^++\tilde{\Gamma}_+-\frac{8\mu_-}{g\tilde{\eta}_+\tilde{\eta}_-}\left(\mathcal{A}_1^+-\mathcal{B}_1^++\frac{4}{g\tilde{\eta}_+\tilde{\eta}_-}\mathcal{C}_1^+\right),\nonumber\\
\psi_3^-&=&\psi_0^-+F_{1,2}\psi_{1,2}^-+\frac{8\mu_-}{g\tilde{\eta}_+\tilde{\eta}_-}\left(\mathcal{A}_2^+-\mathcal{B}_2^++\frac{4}{g\tilde{\eta}_+\tilde{\eta}_-}\mathcal{C}_2^+\right),\nonumber\\
\bar{\psi}_3^-&=&\bar{\psi}_0^-+F_{1,2}\bar{\psi}_{1,2}^-+\frac{8\mu_-}{g\tilde{\eta}_+\tilde{\eta}_-}\left(\mathcal{A}_3^+-\mathcal{B}_3^++\frac{4}{g\tilde{\eta}_+\tilde{\eta}_-}\mathcal{C}_3^+\right),\nonumber
\end{eqnarray}
\begin{eqnarray}
\varphi_3^-&=&\varphi_0^-+\tilde{\Gamma}_-+\frac{8\mu_-}{g\tilde{\eta}_+\tilde{\eta}_-}\left(\mathcal{A}_1^--\mathcal{B}_1^--\frac{4}{g\tilde{\eta}_+\tilde{\eta}_-}\mathcal{C}_1^-\right),\nonumber\\
\psi_3^+&=&\psi_0^++F_{1,2}\psi_{1,2}^+-\frac{8\mu_-}{g\tilde{\eta}_+\tilde{\eta}_-}\left(\mathcal{A}_2^--\mathcal{B}_2^--\frac{4}{g\tilde{\eta}_+\tilde{\eta}_-}\mathcal{C}_2^-\right),\nonumber\\
\bar{\psi}_3^+&=&\bar{\psi}_0^++F_{1,2}\bar{\psi}_{1,2}^+-\frac{8\mu_-}{g\tilde{\eta}_+\tilde{\eta}_-}\left(\mathcal{A}_3^--\mathcal{B}_3^--\frac{4}{g\tilde{\eta}_+\tilde{\eta}_-}\mathcal{C}_3^-\right),\nonumber
\end{eqnarray}
where
\begin{eqnarray}
\tilde{\Gamma}_\pm&=&2\arctan\left[\delta\tan\left(\frac{\varphi_{1,2}^++\varphi_{1,2}^-}{4}\right)\right]\pm2\arctan\left[\delta\tan\left(\frac{\varphi_{1,2}^+-\varphi_{1,2}^-}{4}\right)\right],\nonumber\\
\tilde{\eta}_\pm&=&\mu_+-2\cos\left(\frac{\varphi_{1,2}^+\pm\varphi_{1,2}^-}{2}\right),\nonumber\\
F_{1,2}&=&\frac{\delta}{2}\left[\frac{\sec^2\left(\frac{\varphi_{1,2}^++\varphi_{1,2}^-}{4}\right)}{1+\delta^2\tan^2\left(\frac{\varphi_{1,2}^++\varphi_{1,2}^-}{4}\right)}+\frac{\sec^2\left(\frac{\varphi_{1,2}^+-\varphi_{1,2}^-}{4}\right)}{1+\delta^2\tan^2\left(\frac{\varphi_{1,2}^+-\varphi_{1,2}^-}{4}\right)}\right],\nonumber
\end{eqnarray}

\newpage
\begin{eqnarray}
\mathcal{A}_1^+&=&\cos\left(\frac{\varphi_{1,2}^+}{2}\right)\sin\left(\frac{\varphi_{1,2}^-}{2}\right)\left(\zeta_1^{+(0,1)}\xi_2^{+(0,1)}+\zeta_1^{+(0,2)}\xi_2^{+(0,2)}\right),\nonumber\\
\mathcal{A}_2^+&=&-\cos\left(\frac{\varphi_{1,2}^+}{2}\right)\sin\left(\frac{\varphi_{1,2}^-}{2}\right)\left(\zeta_1^{+(0,1)}p_2^{+(0,1)}+\zeta_1^{+(0,2)}p_2^{+(0,2)}\right)\nonumber\\
&&-\Sigma^+\left(\zeta_1^{+(0,1)}\xi_2^{+(0,1)}+\zeta_1^{+(0,2)}\xi_2^{+(0,2)}\right)\psi_{1,2}^-,\nonumber\\
\mathcal{A}_3^+&=&-\cos\left(\frac{\varphi_{1,2}^+}{2}\right)\sin\left(\frac{\varphi_{1,2}^-}{2}\right)\left(q_2^{+(0,1)}\xi_2^{+(0,1)}+q_2^{+(0,2)}\xi_2^{+(0,2)}\right)\nonumber\\
&&-\Sigma^+\left(\zeta_1^{+(0,1)}\xi_2^{+(0,1)}+\zeta_1^{+(0,2)}\xi_2^{+(0,2)}\right)\bar{\psi}_{1,2}^-,\nonumber
\end{eqnarray}
\begin{eqnarray}
\mathcal{B}_1^+&=&\sin\left(\frac{\varphi_{1,2}^-}{2}\right)\left(\frac{\lambda_2}{\lambda_1}\zeta_1^{+(0,1)}\xi_2^{+(0,2)}+\frac{\lambda_1}{\lambda_2}\zeta_1^{+(0,2)}\xi_2^{+(0,1)}\right),\nonumber\\
\mathcal{B}_2^+&=&-\sin\left(\frac{\varphi_{1,2}^-}{2}\right)\left(\frac{\lambda_2}{\lambda_1}\zeta_1^{+(0,1)}p_2^{+(0,2)}+\frac{\lambda_1}{\lambda_2}\zeta_1^{+(0,2)}p_2^{+(0,1)}\right)\nonumber\\
&&-\Omega^+\left(\frac{\lambda_2}{\lambda_1}\zeta_1^{+(0,1)}\xi_2^{+(0,2)}+\frac{\lambda_1}{\lambda_2}\zeta_1^{+(0,2)}\xi_2^{+(0,1)}\right)\psi_{1,2}^-,\nonumber\\
\mathcal{B}_3^+&=&-\sin\left(\frac{\varphi_{1,2}^-}{2}\right)\left(\frac{\lambda_2}{\lambda_1}q_2^{+(0,1)}\xi_2^{+(0,2)}+\frac{\lambda_1}{\lambda_2}q_2^{+(0,2)}\xi_2^{+(0,1)}\right)\nonumber\\
&&-\Omega^+\left(\frac{\lambda_2}{\lambda_1}\zeta_1^{+(0,1)}\xi_2^{+(0,2)}+\frac{\lambda_1}{\lambda_2}\zeta_1^{+(0,2)}\xi_2^{+(0,1)}\right)\bar{\psi}_{1,2}^-,\nonumber
\end{eqnarray}
\begin{eqnarray}
\mathcal{C}_1^+&=&\sin\left(\frac{\varphi_{1,2}^+}{2}\right)A^+\zeta_1^{+(0,1)}\zeta_1^{+(0,2)}\xi_2^{+(0,1)}\xi_2^{+(0,2)},\nonumber\\
\mathcal{C}_2^+&=&\sin\left(\frac{\varphi_{1,2}^+}{2}\right)A^+\zeta_1^{+(0,1)}\zeta_1^{+(0,2)}\left(\xi_2^{+(0,2)}p_2^{+(0,1)}-\xi_2^{+(0,1)}p_2^{+(0,2)}\right),\nonumber\\
\mathcal{C}_3^+&=&\sin\left(\frac{\varphi_{1,2}^+}{2}\right)A^+\xi_2^{+(0,1)}\xi_2^{+(0,2)}\left(\zeta_1^{+(0,1)}q_2^{+(0,2)}-\zeta_1^{+(0,2)}q_2^{+(0,1)}\right),\nonumber
\end{eqnarray}

\newpage
\begin{eqnarray}
\mathcal{A}_1^-&=&\cos\left(\frac{\varphi_{1,2}^-}{2}\right)\sin\left(\frac{\varphi_{1,2}^+}{2}\right)\left(\zeta_1^{+(0,1)}\xi_2^{+(0,1)}+\zeta_1^{+(0,2)}\xi_2^{+(0,2)}\right),\nonumber\\
\mathcal{A}_2^-&=&-\cos\left(\frac{\varphi_{1,2}^-}{2}\right)\sin\left(\frac{\varphi_{1,2}^+}{2}\right)\left(\partial_zq_1^{+(0,1)}\xi_2^{+(0,1)}+\partial_zq_1^{+(0,2)}\xi_2^{+(0,2)}\right)\nonumber\\
&&-\Sigma^-\left(\zeta_1^{+(0,1)}\xi_2^{+(0,1)}+\zeta_1^{+(0,2)}\xi_2^{+(0,2)}\right)\psi_{1,2}^+,\nonumber\\
\mathcal{A}_3^-&=&\cos\left(\frac{\varphi_{1,2}^-}{2}\right)\sin\left(\frac{\varphi_{1,2}^+}{2}\right)\left(\zeta_1^{+(0,1)}\partial_{\bar{z}}p_1^{+(0,1)}+\zeta_1^{+(0,2)}\partial_{\bar{z}}p_1^{+(0,2)}\right)\nonumber\\
&&-\Sigma^-\left(\zeta_1^{+(0,1)}\xi_2^{+(0,1)}+\zeta_1^{+(0,2)}\xi_2^{+(0,2)}\right)\bar{\psi}_{1,2}^+,\nonumber
\end{eqnarray}
\begin{eqnarray}
\mathcal{B}_1^-&=&\sin\left(\frac{\varphi_{1,2}^+}{2}\right)\left(\frac{\kappa_2}{\kappa_1}\zeta_1^{+(0,1)}\xi_2^{+(0,2)}+\frac{\kappa_1}{\kappa_2}\zeta_1^{+(0,2)}\xi_2^{+(0,1)}\right),\nonumber\\
\mathcal{B}_2^-&=&-\sin\left(\frac{\varphi_{1,2}^+}{2}\right)\left(\frac{\kappa_2}{\kappa_1}\partial_zq_1^{+(0,1)}\xi_2^{+(0,2)}+\frac{\kappa_1}{\kappa_2}\partial_zq_1^{+(0,2)}\xi_2^{+(0,1)}\right)\nonumber\\
&&-\Omega^-\left(\frac{\kappa_2}{\kappa_1}\zeta_1^{+(0,1)}\xi_2^{+(0,2)}+\frac{\kappa_1}{\kappa_2}\zeta_1^{+(0,2)}\xi_2^{+(0,1)}\right)\psi_{1,2}^+,\nonumber\\
\mathcal{B}_3^-&=&\sin\left(\frac{\varphi_{1,2}^+}{2}\right)\left(\frac{\kappa_2}{\kappa_1}\zeta_1^{+(0,1)}\partial_{\bar{z}}p_1^{+(0,2)}+\frac{\kappa_1}{\kappa_2}\zeta_1^{+(0,2)}\partial_{\bar{z}}p_1^{+(0,1)}\right)\nonumber\\
&&-\Omega^-\left(\frac{\kappa_2}{\kappa_1}\zeta_1^{+(0,1)}\xi_2^{+(0,2)}+\frac{\kappa_1}{\kappa_2}\zeta_1^{+(0,2)}\xi_2^{+(0,1)}\right)\bar{\psi}_{1,2}^+,\nonumber
\end{eqnarray}
\begin{eqnarray}
\mathcal{C}_1^-&=&\sin\left(\frac{\varphi_{1,2}^-}{2}\right)A^-\zeta_1^{+(0,1)}\zeta_1^{+(0,2)}\xi_2^{+(0,1)}\xi_2^{+(0,2)},\nonumber\\
\mathcal{C}_2^-&=&-\sin\left(\frac{\varphi_{1,2}^-}{2}\right)A^-\xi_2^{+(0,1)}\xi_2^{+(0,2)}\left(\zeta_1^{+(0,2)}\partial_zq_1^{+(0,1)}-\zeta_1^{+(0,1)}\partial_zq_1^{+(0,2)}\right),\nonumber\\
\mathcal{C}_3^-&=&-\sin\left(\frac{\varphi_{1,2}^-}{2}\right)A^-\zeta_1^{+(0,1)}\zeta_1^{+(0,2)}\left(\xi_2^{+(0,2)}\partial_{\bar{z}}p_1^{+(0,1)}-\xi_2^{+(0,1)}\partial_{\bar{z}}p_1^{+(0,2)}\right),\nonumber
\end{eqnarray}

\newpage
\begin{eqnarray}
\Sigma^+&=&\cos\left(\frac{\varphi_{1,2}^+}{2}\right)\Omega^+-\frac{1}{2}\sin\left(\frac{\varphi_{1,2}^-}{2}\right)\sin\left(\frac{\varphi_{1,2}^+}{2}\right),\nonumber\\
\Omega^+&=&-\sin\left(\frac{\varphi_{1,2}^-}{2}\right)\left[\frac{1}{\tilde{\eta}_+}\sin\left(\frac{\varphi_{1,2}^++\varphi_{1,2}^-}{2}\right)+\frac{1}{\tilde{\eta}_-}\sin\left(\frac{\varphi_{1,2}^+-\varphi_{1,2}^-}{2}\right)\right],\nonumber\\
A^+&=&\cos\left(\frac{\varphi_{1,2}^-}{2}\right)\left(a+\cos\varphi_{1,2}^+-\cos\varphi_{1,2}^-\right)-2\mu_+\cos\left(\frac{\varphi_{1,2}^+}{2}\right),\nonumber
\end{eqnarray}
\begin{eqnarray}
\Sigma^-&=&\cos\left(\frac{\varphi_{1,2}^-}{2}\right)\Omega^--\frac{1}{2}\sin\left(\frac{\varphi_{1,2}^-}{2}\right)\sin\left(\frac{\varphi_{1,2}^+}{2}\right),\nonumber\\
\Omega^-&=&-\sin\left(\frac{\varphi_{1,2}^+}{2}\right)\left[\frac{1}{\tilde{\eta}_+}\sin\left(\frac{\varphi_{1,2}^++\varphi_{1,2}^-}{2}\right)-\frac{1}{\tilde{\eta}_-}\sin\left(\frac{\varphi_{1,2}^+-\varphi_{1,2}^-}{2}\right)\right],\nonumber\\
A^-&=&\cos\left(\frac{\varphi_{1,2}^+}{2}\right)\left(a-\cos\varphi_{1,2}^++\cos\varphi_{1,2}^-\right)-2\mu_+\cos\left(\frac{\varphi_{1,2}^-}{2}\right),\nonumber
\end{eqnarray}
and denoted
\begin{eqnarray}
\varphi_{1,2}^\pm=\varphi_1^\pm-\varphi_2^\pm,\qquad \psi_{1,2}^\pm=\psi_1^\pm-\psi_2^\pm,\qquad \bar{\psi}_{1,2}^\pm=\bar{\psi}_1^\pm-\bar{\psi}_2^\pm.\nonumber
\end{eqnarray}
From the Backlund eqns. we get (see app. A)
\begin{eqnarray}
\zeta_1^{+(0,k)}=-\frac{\kappa_k}{8}\frac{(\psi_0^--\psi_k^-)}{\cos\left(\frac{\varphi_0^-+\varphi_k^-}{2}\right)},&&\qquad
\xi_2^{+(0,k)}=\frac{1}{\kappa_k}\frac{(\bar{\psi}_0^-+\bar{\psi}_k^-)}{\cos\left(\frac{\varphi_0^--\varphi_k^-}{2}\right)},\nonumber\\
\partial_zq_1^{+(0,k)}=\frac{\lambda_k g}{4}\sin\left(\frac{\varphi_0^++\varphi_k^+}{2}\right),&&\qquad
p_2^{+(0,k)}=\frac{2 g}{\kappa_k}\sin\left(\frac{\varphi_0^-+\varphi_k^-}{2}\right),\nonumber\\
q_2^{+(0,k)}=-\frac{\kappa_k g}{4}\sin\left(\frac{\varphi_0^--\varphi_k^-}{2}\right),&&\qquad
\partial_{\bar{z}}p_1^{+(0,k)}=\frac{2 g}{\lambda_k}\sin\left(\frac{\varphi_0^+-\varphi_k^+}{2}\right).\nonumber
\end{eqnarray}

\section*{4\hspace{0.5cm}$1$-Soliton Solution}
Setting $\phi_0^\pm=0$ in the Backlund eqns.  (\ref{eq1})-(\ref{p4}) we find in components,
\begin{eqnarray}
\partial_{\bar{z}}\zeta_1^{+(0,1)}&=&-\frac{g^2}{\gamma_1}\cos\left(\frac{\varphi_1^+}{2}\right)\cos\left(\frac{\varphi_1^-}{2}\right)\zeta_1^{+(0,1)},\nonumber\\ \partial_{z}\zeta_1^{+(0,1)}&=&\gamma_1\cos\left(\frac{\varphi_1^+}{2}\right)\cos\left(\frac{\varphi_1^-}{2}\right)\zeta_1^{+(0,1)},\nonumber\\
\partial_{\bar{z}}\xi_2^{+(0,1)}&=&-\frac{g^2}{\gamma_1}\cos\left(\frac{\varphi_1^+}{2}\right)\cos\left(\frac{\varphi_1^-}{2}\right)\xi_2^{+(0,1)},\nonumber\\ \partial_{z}\xi_2^{+(0,1)}&=&\gamma_1\cos\left(\frac{\varphi_1^+}{2}\right)\cos\left(\frac{\varphi_1^-}{2}\right)\xi_2^{+(0,1)},\nonumber\\
\partial_{\bar{z}}\varphi_1^\pm&=&-\frac{2g^2}{\gamma_1}\sin\left(\frac{\varphi_1^\pm}{2}\right)\cos\left(\frac{\varphi_1^\mp}{2}\right), \nonumber\\ \partial_{z}\varphi_1^\pm&=&2\gamma_1\sin\left(\frac{\varphi_1^\pm}{2}\right)\cos\left(\frac{\varphi_1^\mp}{2}\right).\nonumber
\end{eqnarray}
Integrating the above eqns. we get the 1-soliton solution, 
\begin{eqnarray}
\psi_1^-&=&\frac{8}{\kappa_1}\cos\left(\frac{\varphi_1^-}{2}\right)\zeta_1^{+(0,1)},\qquad \psi_1^+=-\lambda_1\cos\left(\frac{\varphi_1^+}{2}\right)\xi_2^{+(0,1)},\nonumber\\
\bar{\psi}_1^-&=&\kappa_1\cos\left(\frac{\varphi_1^-}{2}\right)\xi_2^{+(0,1)},\qquad \bar{\psi}_1^+=-\frac{8}{\lambda_1}\cos\left(\frac{\varphi_1^+}{2}\right)\zeta_1^{+(0,1)},\nonumber
\end{eqnarray} 
\begin{eqnarray}
\varphi_1^\pm&=&2\arctan(a_1\rho_1)\pm2\arctan(b_1\rho_1),\nonumber\\
\zeta_1^{+(0,1)}&=&\xi_2^{+(0,1)}=\epsilon_1\chi_1,\nonumber\\ \chi_1&=&\frac{\rho_1}{\sqrt{(1+a_1^2\rho_1^2)(1+b_1^2\rho_1^2)}},\nonumber
\end{eqnarray}
where $a_1$ and $b_1$ are arbitrary constants,  $\epsilon_1$ is a grassmann parameter and 
\begin{eqnarray}
\rho_1=\exp\left(\gamma_1z-\frac{g^2}{\gamma_1}\bar{z}\right).\nonumber
\end{eqnarray} 
The 1-soliton solution constructed in this section can be obtained from those  of \cite{n2} by  relating parameters since they both involve a single grassmann parameter.

\section*{5\hspace{0.5cm}$2$-Soliton Solution}

For the  $2$-soliton case  we obtain from the superposition formulae (\ref{ansatz}) 
\begin{eqnarray}
\varphi_3^+&=&\varphi_3^{+(0)}+\varphi_3^{+(1)}\epsilon_1\epsilon_2,\nonumber\\
\varphi_3^{+(0)}&=&2\arctan\left[\delta\tan\left(\frac{\varphi_{1,2}^++\varphi_{1,2}^-}{4}\right)\right]+2\arctan\left[\delta\tan\left(\frac{\varphi_{1,2}^+-\varphi_{1,2}^-}{4}\right)\right],\nonumber\\
\varphi_3^{+(1)}&=&\frac{8\mu_-}{g\tilde{\eta}_+\tilde{\eta}_-}\sin\left(\frac{\varphi_{1,2}^-}{2}\right)\left(\frac{\lambda_2}{\lambda_1}-\frac{\lambda_1}{\lambda_2}\right)\chi_1\chi_2,\nonumber
\end{eqnarray}
\begin{eqnarray}
\psi_3^-&=&\epsilon_1\psi_3^{-(1)}+\epsilon_2\psi_3^{-(2)},\nonumber\\
\psi_3^{-(1)}&=&\frac{8}{\kappa_1}F_{1,2}\cos\left(\frac{\varphi_1^-}{2}\right)\chi_1\nonumber\\
&&+\frac{16}{\kappa_1\gamma_1}\frac{\mu_-}{\tilde{\eta}_+\tilde{\eta}_-}\sin\left(\frac{\varphi_{1,2}^-}{2}\right)\chi_1\left[\gamma_2\sin\left(\frac{\varphi_2^-}{2}\right)-\gamma_1\cos\left(\frac{\varphi_{1,2}^+}{2}\right)\sin\left(\frac{\varphi_1^-}{2}\right)\right],\nonumber\\
\psi_3^{-(2)}&=&-\frac{8}{\kappa_2}F_{1,2}\cos\left(\frac{\varphi_2^-}{2}\right)\chi_2\nonumber\\
&&+\frac{16}{\kappa_2\gamma_2}\frac{\mu_-}{\tilde{\eta}_+\tilde{\eta}_-}\sin\left(\frac{\varphi_{1,2}^-}{2}\right)\chi_2\left[\gamma_1\sin\left(\frac{\varphi_1^-}{2}\right)-\gamma_2\cos\left(\frac{\varphi_{1,2}^+}{2}\right)\sin\left(\frac{\varphi_2^-}{2}\right)\right],\nonumber
\end{eqnarray}
\begin{eqnarray}
\bar{\psi}_3^-&=&\epsilon_1\bar{\psi}_3^{-(1)}+\epsilon_2\bar{\psi}_3^{-(2)},\nonumber\\
\bar{\psi}_3^{-(1)}&=&\kappa_1F_{1,2}\cos\left(\frac{\varphi_1^-}{2}\right)\chi_1\nonumber\\
&&+\frac{2\kappa_1}{\gamma_2}\frac{\mu_-}{\tilde{\eta}_+\tilde{\eta}_-}\sin\left(\frac{\varphi_{1,2}^-}{2}\right)\chi_1\left[\gamma_1\sin\left(\frac{\varphi_2^-}{2}\right)-\gamma_2\cos\left(\frac{\varphi_{1,2}^+}{2}\right)\sin\left(\frac{\varphi_1^-}{2}\right)\right],\nonumber\\
\bar{\psi}_3^{-(2)}&=&-\kappa_2F_{1,2}\cos\left(\frac{\varphi_2^-}{2}\right)\chi_2\nonumber\\
&&+\frac{2\kappa_2}{\gamma_1}\frac{\mu_-}{\tilde{\eta}_+\tilde{\eta}_-}\sin\left(\frac{\varphi_{1,2}^-}{2}\right)\chi_2\left[\gamma_2\sin\left(\frac{\varphi_1^-}{2}\right)-\gamma_1\cos\left(\frac{\varphi_{1,2}^+}{2}\right)\sin\left(\frac{\varphi_2^-}{2}\right)\right],\nonumber
\end{eqnarray}


\begin{eqnarray}
\varphi_3^-&=&\varphi_3^{-(0)}+\varphi_3^{-(1)}\epsilon_1\epsilon_2,\nonumber\\
\varphi_3^{-(0)}&=&2\arctan\left[\delta\tan\left(\frac{\varphi_{1,2}^++\varphi_{1,2}^-}{4}\right)\right]-2\arctan\left[\delta\tan\left(\frac{\varphi_{1,2}^+-\varphi_{1,2}^-}{4}\right)\right],\nonumber\\
\varphi_3^{-(1)}&=&-\frac{8\mu_-}{g\tilde{\eta}_+\tilde{\eta}_-}\sin\left(\frac{\varphi_{1,2}^+}{2}\right)\left(\frac{\kappa_2}{\kappa_1}-\frac{\kappa_1}{\kappa_2}\right)\chi_1\chi_2,\nonumber
\end{eqnarray}
\begin{eqnarray}
\psi_3^+&=&\epsilon_1\psi_3^{+(1)}+\epsilon_2\psi_3^{+(2)},\nonumber\\
\psi_3^{+(1)}&=&-\lambda_1F_{1,2}\cos\left(\frac{\varphi_1^+}{2}\right)\chi_1\nonumber\\
&&-\frac{2\lambda_1}{\gamma_1}\frac{\mu_-}{\tilde{\eta}_+\tilde{\eta}_-}\sin\left(\frac{\varphi_{1,2}^+}{2}\right)\chi_1\left[\gamma_2\sin\left(\frac{\varphi_2^+}{2}\right)-\gamma_1\cos\left(\frac{\varphi_{1,2}^-}{2}\right)\sin\left(\frac{\varphi_1^+}{2}\right)\right],\nonumber\\
\psi_3^{+(2)}&=&\lambda_2F_{1,2}\cos\left(\frac{\varphi_2^+}{2}\right)\chi_2\nonumber\\
&&-\frac{2\lambda_2}{\gamma_2}\frac{\mu_-}{\tilde{\eta}_+\tilde{\eta}_-}\sin\left(\frac{\varphi_{1,2}^+}{2}\right)\chi_2\left[\gamma_1\sin\left(\frac{\varphi_1^+}{2}\right)-\gamma_2\cos\left(\frac{\varphi_{1,2}^-}{2}\right)\sin\left(\frac{\varphi_2^+}{2}\right)\right],\nonumber
\end{eqnarray}
\begin{eqnarray}
\bar{\psi}_3^+&=&\epsilon_1\bar{\psi}_3^{+(1)}+\epsilon_2\bar{\psi}_3^{+(2)},\nonumber\\
\bar{\psi}_3^{+(1)}&=&-\frac{8}{\lambda_1}F_{1,2}\cos\left(\frac{\varphi_1^+}{2}\right)\chi_1\nonumber\\
&&-\frac{16}{\lambda_1\gamma_2}\frac{\mu_-}{\tilde{\eta}_+\tilde{\eta}_-}\sin\left(\frac{\varphi_{1,2}^+}{2}\right)\chi_1\left[\gamma_1\sin\left(\frac{\varphi_2^+}{2}\right)-\gamma_2\cos\left(\frac{\varphi_{1,2}^-}{2}\right)\sin\left(\frac{\varphi_1^+}{2}\right)\right],\nonumber\\
\bar{\psi}_3^{+(2)}&=&\frac{8}{\lambda_2}F_{1,2}\cos\left(\frac{\varphi_2^+}{2}\right)\chi_2\nonumber\\
&&-\frac{16}{\lambda_2\gamma_1}\frac{\mu_-}{\tilde{\eta}_+\tilde{\eta}_-}\sin\left(\frac{\varphi_{1,2}^+}{2}\right)\chi_2\left[\gamma_2\sin\left(\frac{\varphi_1^+}{2}\right)-\gamma_1\cos\left(\frac{\varphi_{1,2}^-}{2}\right)\sin\left(\frac{\varphi_2^+}{2}\right)\right],\nonumber
\end{eqnarray}
\newpage

where
\begin{eqnarray}
\varphi_k^\pm&=&2\arctan(a_k\rho_k)\pm2\arctan(b_k\rho_k),\nonumber\\
\chi_k&=&\frac{\rho_k}{\sqrt{(1+a_k^2\rho_k^2)(1+b_k^2\rho_k^2)}},\nonumber
\end{eqnarray}
$k=1,2$, $a_k$ and $b_k$ are arbitrary constants,  $\epsilon_k$ is a grassmann constant and
\begin{eqnarray}
\rho_k=\exp\left(\gamma_kz-\frac{g^2}{\gamma_k}\bar{z}\right).\nonumber
\end{eqnarray}
Notice that the 2-soliton solution constructed in this section generalizes  those constructed in ref. \cite{n2}) involving a single  grassmann parameter.  

Both  1- and 2-soliton solutions presented above were verified to satisfy the equations of motion.
\vskip 1cm
 \noindent
{\bf Acknowledgments} \\
\vskip .1cm \noindent
{  LHY acknowledges support from Fapesp, JFG and AHZ thank CNPq for partial support.}
\bigskip

\section*{Appendix A}

In order to simplify notation  let us introduce $\varphi_{\pm}^{(-)}=\varphi_{1}^{-}\pm \varphi_{2}^{-}$,
$\varphi_{\pm}^{(+)}=\varphi_{1}^{+}\pm \varphi_{2}^{+}$ and similar notation for the other fields.

In components  eqn. ({\ref{F1F2}) becomes
\begin{eqnarray}
\begin{array}{l}
\bullet\,\,\bar{D}_{+}\mathcal{F}=-\frac{\kappa g}{4}\sin\left(\frac{\phi_{1}^{-}-\phi_{2}^{-}}{2}\right),\nonumber\\
\qquad \qquad \Downarrow\nonumber\\
q_{2}^{+}=-\frac{\kappa g}{4}\sin\left(\frac{\varphi_{-}^{(-)}}{2}\right), \qquad
\partial_{\bar{z}}\zeta_{1}^{+}=-\frac{\kappa g}{8}\cos\left(\frac{\varphi_{-}^{(-)}}{2}\right)\bar{\psi}_{-}^{(+)},\qquad \partial_{z}\zeta_{2}^{+}=\frac{\kappa g}{8}\cos\left(\frac{\varphi_{-}^{(-)}}{2}\right)\psi_{-}^{(+)},\nonumber\\
\partial_{\bar{z}}\partial_{z}q_{1}^{+}=\frac{\kappa g}{8}\cos\left(\frac{\varphi_{-}^{(-)}}{2}\right)F_{-}^{(-)}+\frac{\kappa g}{16}\sin\left(\frac{\varphi_{-}^{(-)}}{2}\right)\psi_{-}^{(+)}\bar{\psi}_{-}^{(+)}.\nonumber
\end{array}
\end{eqnarray}

\begin{eqnarray}
\begin{array}{l}
\bullet\,\,D_{+}\mathcal{G}=-\frac{2g}{\kappa}\sin\left(\frac{\phi_{1}^{-}+\phi_{2}^{-}}{2}\right),\nonumber\\
\qquad \qquad \Downarrow\nonumber\\
p_{2}^{+}=\frac{2g}{\kappa}\sin\left(\frac{\varphi_{+}^{(-)}}{2}\right), \qquad
\partial_{\bar{z}}\xi_{1}^{+}=\frac{g}{\kappa}\cos\left(\frac{\varphi_{+}^{(-)}}{2}\right)\bar{\psi}_{+}^{(+)},\qquad
\partial_{z}\xi_{2}^{+}=-\frac{g}{\kappa}\cos\left(\frac{\varphi_{+}^{(-)}}{2}\right)\psi_{+}^{(+)},\nonumber\\
\partial_{\bar{z}}\partial_{z}p_{1}^{+}=-\frac{g}{\kappa}\cos\left(\frac{\varphi_{+}^{(-)}}{2}\right)F_{+}^{(-)}-\frac{g}{2\kappa}\sin\left(\frac{\varphi_{+}^{(-)}}{2}\right)\psi_{+}^{(+)}\bar{\psi}_{+}^{(+)}.\nonumber
\end{array}
\end{eqnarray}

\pagebreak 
Similarly we find for (\ref{G1G2}), 
\begin{eqnarray}
\begin{array}{l}
\bullet\,\,\bar{D}_{-}\mathcal{G}=\frac{2g}{\lambda}\sin\left(\frac{\phi_{1}^{+}-\phi_{2}^{+}}{2}\right),\nonumber\\
\qquad \qquad \Downarrow\nonumber\\
q_{2}^{-}=\frac{2g}{\lambda}\sin\left(\frac{\varphi_{-}^{(+)}}{2}\right),\qquad
\partial_{\bar{z}}\zeta_{1}^{-}=\frac{g}{\lambda}\cos\left(\frac{\varphi_{-}^{(+)}}{2}\right)\bar{\psi}_{-}^{(-)},\qquad
\partial_{z}\zeta_{2}^{-}=-\frac{g}{\lambda}\cos\left(\frac{\varphi_{-}^{(+)}}{2}\right)\psi_{-}^{(-)},\nonumber\\
\partial_{\bar{z}}\partial_{z}q_{1}^{-}=-\frac{g}{\lambda}\cos\left(\frac{\varphi_{-}^{(+)}}{2}\right)F_{-}^{(+)}-\frac{g}{2\lambda}\sin\left(\frac{\varphi_{-}^{(+)}}{2}\right)\psi_{-}^{(-)}\bar{\psi}_{-}^{(-)}.\nonumber
\end{array}
\end{eqnarray}

\begin{eqnarray}
\begin{array}{l}
\bullet\,\,D_{-}\mathcal{F}=\frac{\lambda g}{4}\sin\left(\frac{\phi_{1}^{+}+\phi_{2}^{+}}{2}\right),\nonumber\\
\qquad \qquad \Downarrow\nonumber\\
p_{2}^{-}=-\frac{\lambda g}{4}\sin\left(\frac{\varphi_{+}^{(+)}}{2}\right),\qquad
\partial_{\bar{z}}\xi_{1}^{-}=-\frac{\lambda g}{8}\cos\left(\frac{\varphi_{+}^{(+)}}{2}\right)\bar{\psi}_{+}^{(-)},\qquad
\partial_{z}\xi_{2}^{-}=\frac{\lambda g}{8}\cos\left(\frac{\varphi_{+}^{(+)}}{2}\right)\psi_{+}^{(-)},\nonumber\\
\partial_{\bar{z}}\partial_{z}p_{1}^{-}=\frac{\lambda g}{8}\cos\left(\frac{\varphi_{+}^{(+)}}{2}\right)F_{+}^{(+)}+\frac{\lambda g}{16}\sin\left(\frac{\varphi_{+}^{(+)}}{2}\right)\psi_{+}^{(-)}\bar{\psi}_{+}^{(-)}.\nonumber
\end{array}
\end{eqnarray}

From (\ref{f1}) and (\ref{f2}),
\begin{eqnarray}
\begin{array}{l}
\bullet\,\,D_{+}\phi_{1}^{+}=D_{+}\phi_{2}^{+}-\frac{8}{\kappa}\mathcal{F}\cos\left(\frac{\phi_{1}^{-}+\phi_{2}^{-}}{2}\right),\nonumber\\
\qquad \qquad \Downarrow\nonumber\\
\psi_{-}^{(-)}=-\frac{8}{\kappa}\zeta_{1}^{+}\cos\left(\frac{\varphi_{+}^{(-)}}{2}\right),\qquad F_{-}^{(+)}=-\frac{8}{\kappa}q_{2}^{+}\cos\left(\frac{\varphi_{+}^{(-)}}{2}\right),\nonumber\\
\partial_{z}\varphi_{-}^{(+)}=-\frac{4}{\kappa}\sin\left(\frac{\varphi_{+}^{(-)}}{2}\right)\zeta_{1}^{+}\psi_{+}^{(+)}-\frac{8}{\kappa}\partial_{z}q_{1}^{+}\cos\left(\frac{\varphi_{+}^{(-)}}{2}\right).\nonumber
\end{array}
\end{eqnarray}

\begin{eqnarray}
\begin{array}{l}
\bullet\,\,\bar{D}_{+}\phi_{1}^{+}=-\bar{D}_{+}\phi_{2}^{+}+\kappa\mathcal{G}\cos\left(\frac{\phi_{1}^{-}-\phi_{2}^{-}}{2}\right),\nonumber\\
\qquad \qquad \Downarrow\nonumber\\
\bar{\psi}_{+}^{(-)}=\kappa\xi_{2}^{+}\cos\left(\frac{\varphi_{-}^{(-)}}{2}\right),\qquad F_{+}^{(+)}=\kappa p_{2}^{+}\cos\left(\frac{\varphi_{-}^{(-)}}{2}\right),\nonumber\\
\partial_{\bar{z}}\varphi_{+}^{(+)}=\frac{\kappa}{2}\sin\left(\frac{\varphi_{-}^{(-)}}{2}\right)\xi_{2}^{+}\bar{\psi}_{-}^{(+)}+\kappa\partial_{\bar{z}}p_{1}^{+}\cos\left(\frac{\varphi_{-}^{(-)}}{2}\right).\nonumber
\end{array}
\end{eqnarray}

From (\ref{g1}) and (\ref{g2}),

\begin{eqnarray}
\begin{array}{l}
\bullet\,\,D_{-}\phi_{1}^{-}=D_{-}\phi_{2}^{-}+\lambda\mathcal{G}\cos\left(\frac{\phi_{1}^{+}+\phi_{2}^{+}}{2}\right),\nonumber\\
\qquad \qquad \Downarrow\nonumber\\
\psi_{-}^{(+)}=\lambda\zeta_{1}^{-}\cos\left(\frac{\varphi_{+}^{(+)}}{2}\right),\qquad F_{-}^{(-)}=\lambda q_{2}^{-}\cos\left(\frac{\varphi_{+}^{(+)}}{2}\right),\nonumber\\
\partial_{z}\varphi_{-}^{(-)}=\frac{\lambda}{2}\sin\left(\frac{\varphi_{+}^{(+)}}{2}\right)\zeta_{1}^{-}\psi_{+}^{(-)}+\lambda\partial_{z}q_{1}^{-}\cos\left(\frac{\varphi_{+}^{(+)}}{2}\right).\nonumber
\end{array}
\end{eqnarray}

\begin{eqnarray}
\begin{array}{l}
\bullet\,\,\bar{D}_{-}\phi_{1}^{-}=-\bar{D}_{-}\phi_{2}^{-}-\frac{8}{\lambda}\mathcal{F}\cos\left(\frac{\phi_{1}^{+}-\phi_{2}^{+}}{2}\right),\nonumber\\
\qquad \qquad \Downarrow\nonumber\\
\bar{\psi}_{+}^{(+)}=-\frac{8}{\lambda}\xi_{2}^{-}\cos\left(\frac{\varphi_{-}^{(+)}}{2}\right),\qquad F_{+}^{(-)}=-\frac{8}{\lambda}p_{2}^{-}\cos\left(\frac{\varphi_{-}^{(+)}}{2}\right),\nonumber\\
\partial_{\bar{z}}\varphi_{+}^{(-)}=-\frac{4}{\lambda}\sin\left(\frac{\varphi_{-}^{(+)}}{2}\right)\xi_{2}^{-}\bar{\psi}_{-}^{(-)}-\frac{8}{\lambda}\partial_{\bar{z}}p_{1}^{-}\cos\left(\frac{\varphi_{-}^{(+)}}{2}\right).\nonumber
\end{array}
\end{eqnarray}

\section*{Appendix B}

Applying the permutability theorem to eqns. (\ref{equation1}) and  (\ref{equation3}) after neglecting the contribution proportional to fermionic superfields, we obtain the following relations 
\begin{eqnarray}
\gamma_1s_{0,1}^+c_{0,1}^-+\gamma_2s_{1,3}^+c_{1,3}^-&=&\gamma_2s_{0,2}^+c_{0,2}^-+\gamma_1s_{2,3}^+c_{2,3}^-,\nonumber\\
\gamma_1s_{0,1}^-c_{0,1}^++\gamma_2s_{1,3}^-c_{1,3}^+&=&\gamma_2s_{0,2}^-c_{0,2}^++\gamma_1s_{2,3}^-c_{2,3}^+.\nonumber
\end{eqnarray}
Summing and subtracting the above eqns., we find
\begin{eqnarray}
&&\gamma_1\left[(s_{0,1}^+c_{0,1}^-\pm s_{0,1}^-c_{0,1}^+)-(s_{2,3}^+c_{2,3}^-\pm s_{2,3}^-c_{2,3}^+)\right]\nonumber\\
&&+\gamma_2\left[(s_{1,3}^+c_{1,3}^-\pm s_{1,3}^-c_{1,3}^+)-(s_{0,2}^+c_{0,2}^-\pm s_{0,2}^-c_{0,2}^+)\right]=0.\label{p1}
\end{eqnarray}
Using the identity
\begin{eqnarray}\label{id}
\sin a\cos b\pm \sin b\cos a=\sin\left(a\pm b\right),
\end{eqnarray} 
and eqns. (\ref{x1}) and (\ref{x2}) we can rewrite (\ref{p1}) as 
\begin{eqnarray}
&&\gamma_1\left\{\sin\left[\left(\frac{\phi_0^++\phi_1^+}{2}\right)\pm\left(\frac{\phi_0^-+\phi_1^-}{2}\right)\right]-\sin\left[\left(\frac{\phi_2^++\phi_3^+}{2}\right)\pm\left(\frac{\phi_2^-+\phi_3^-}{2}\right)\right]\right\}\nonumber\\
&&\gamma_2\left\{\sin\left[\left(\frac{\phi_1^++\phi_3^+}{2}\right)\pm\left(\frac{\phi_1^-+\phi_3^-}{2}\right)\right]-\sin\left[\left(\frac{\phi_0^++\phi_2^+}{2}\right)\pm\left(\frac{\phi_0^-+\phi_2^-}{2}\right)\right]\right\}=0.\nonumber
\end{eqnarray} 
Using the fact that 
\begin{eqnarray}
\sin a-\sin b=2\cos\left(\frac{a+b}{2}\right)\sin\left(\frac{a-b}{2}\right),\nonumber
\end{eqnarray}
yields
\begin{eqnarray}
2\cos\left(Y^+\pm Y^-\right)\left\{
\gamma_1\sin\left[(X_{1,2}^+\pm X_{1,2}^-)-(X_{3,0}^+\pm X_{3,0}^-) \right]\right.\nonumber\\
\left.+\gamma_2\sin\left[(X_{1,2}^+\pm X_{1,2}^-)+(X_{3,0}^+\pm X_{3,0}^-) \right]\right\}=0,\nonumber
\end{eqnarray}
where we have denoted
\begin{eqnarray}
Y^{\pm}&=&\frac{\phi_0^\pm+\phi_1^\pm+\phi_2^\pm+\phi_3^\pm}{4},\nonumber\\
X_{j,k}^\pm&=&\frac{\phi_j^\pm-\phi_k^\pm}{4}.\nonumber
\end{eqnarray}
from where it follows that 
\begin{eqnarray}
(\gamma_1+\gamma_2)\sin\left(X_{1,2}^+\pm X_{1,2}^-\right)\cos\left(X_{3,0}^+\pm X_{3,0}^-\right)=\nonumber\\
(\gamma_1-\gamma_2)\sin\left(X_{3,0}^+\pm X_{3,0}^-\right)\cos\left(X_{1,2}^+\pm X_{1,2}^-\right),\nonumber
\end{eqnarray}
or,
\begin{eqnarray}
\tan\left(X_{3,0}^+\pm X_{3,0}^-\right)=\left(\frac{\gamma_1+\gamma_2}{\gamma_1-\gamma_2}\right)\tan\left(X_{1,2}^+\pm X_{1,2}^-\right),\nonumber
\end{eqnarray}
and therefore
\begin{eqnarray}
\left(\frac{\phi_3^+-\phi_0^+}{4}\right)\pm\left(\frac{\phi_3^--\phi_0^-}{4}\right)=\arctan\left[\delta\tan\left(X_{1,2}^+\pm X_{1,2}^-\right)\right],\nonumber
\end{eqnarray}
where $\delta=\left(\frac{\gamma_1+\gamma_2}{\gamma_1-\gamma_2}\right)$. 
Adding and subtracting  the above expressions we obtain
\begin{eqnarray}
\phi_3^\pm=\phi_0^\pm+\Gamma_{\pm},\nonumber
\end{eqnarray}
with
\begin{eqnarray}
\Gamma_\pm=2\arctan\left[\delta\tan\left(X_{1,2}^++ X_{1,2}^-\right)\right]\pm2\arctan\left[\delta\tan\left(X_{1,2}^+- X_{1,2}^-\right)\right].\nonumber
\end{eqnarray}

\section*{Appendix C}

Relations (\ref{relation1})and (\ref{relation2}) can be  written in matrix form,
\begin{eqnarray}\label{matrix}
\left(\begin{array}{c}\mathcal{F}^{(1,3)}\\\mathcal{F}^{(2,3)}\end{array}\right)=\frac{1}{Z}\left(\begin{array}{cc}A&-B\\C&-D\end{array}\right)\left(\begin{array}{c}\mathcal{F}^{(0,1)}\\\mathcal{F}^{(0,2)}\end{array}\right)
\end{eqnarray}
where
\begin{eqnarray}
A&=&\kappa_2\lambda_2(\bar{c}_{0,1}^+c_{2,3}^-+\bar{c}_{2,3}^+c_{0,1}^-),\nonumber\\
B&=&\kappa_2\lambda_1\bar{c}_{0,2}^+c_{2,3}^-+\kappa_1\lambda_2\bar{c}_{2,3}^+c_{0,2}^-,\nonumber\\
C&=&\kappa_2\lambda_1\bar{c}_{1,3}^+c_{0,1}^-+\kappa_1\lambda_2\bar{c}_{0,1}^+c_{1,3}^-,\nonumber\\
D&=&\kappa_1\lambda_1(\bar{c}_{0,2}^+c_{1,3}^-+\bar{c}_{1,3}^+c_{0,2}^-),\nonumber\\
Z&=&\kappa_2\lambda_1\bar{c}_{1,3}^+c_{2,3}^--\kappa_1\lambda_2\bar{c}_{2,3}^+c_{1,3}^-.\label{abcd}
\end{eqnarray}
Introduce eqn. (\ref{ansatz}) into  expressions (\ref{abcd}).  Consider now the following expansions
\begin{eqnarray}
c_{k,3}^-&=&c_{k,\Gamma_-}\left(1-\frac{\Delta_-^2}{8}\right)-\frac{\Delta_-}{2}s_{k,\Gamma_-},\nonumber\\
\bar{c}_{k,3}^+&=&\bar{c}_{k,\Gamma_-}\left(1-\frac{\Delta_+^2}{8}\right)+\frac{\Delta_+}{2}\bar{s}_{k,\Gamma_+},\nonumber
\end{eqnarray}
where we have denoted
\begin{eqnarray}
c_{k,\Gamma_-}&=&\cos\left(\frac{\phi_k^{-}+\phi_0^{-}+\Gamma_-}{2}\right)=c_{k,0}^-\sigma_+-s_{k,0}^-\rho_-,\nonumber\\
s_{k,\Gamma_-}&=&\sin\left(\frac{\phi_k^{-}+\phi_0^{-}+\Gamma_-}{2}\right)=s_{k,0}^-\sigma_++c_{k,0}^-\rho_-,\nonumber\\
\bar{c}_{k,\Gamma_+}&=&\cos\left(\frac{\phi_k^{+}-\phi_0^{+}-\Gamma_+}{2}\right)=\bar{c}_{k,0}^+\sigma_-+\bar{s}_{k,0}^+\rho_+\nonumber\\
\bar{s}_{k,\Gamma_+}&=&\sin\left(\frac{\phi_k^{+}-\phi_0^{+}-\Gamma_+}{2}\right)=\bar{s}_{k,0}^+\sigma_--\bar{c}_{k,0}^+\rho_+,\nonumber
\end{eqnarray}
and
\begin{eqnarray}
\sigma_{\pm}&=&\frac{1\pm\delta^2\tan\left(\frac{x+y}{4}\right)\tan\left(\frac{x-y}{4}\right)}{\sqrt{1+\delta^2\tan^2\left(\frac{x+y}{4}\right)}\sqrt{1+\delta^2\tan^2\left(\frac{x-y}{4}\right)}},\nonumber\\
\rho_{\pm}&=&\frac{\delta\left[\tan\left(\frac{x+y}{4}\right)\pm\tan\left(\frac{x-y}{4}\right)\right]}{\sqrt{1+\delta^2\tan^2\left(\frac{x+y}{4}\right)}\sqrt{1+\delta^2\tan^2\left(\frac{x-y}{4}\right)}}.\nonumber
\end{eqnarray}
Next, we expand  the expressions for $A$, $B$, $C$, $D$  and  $Z$ in power series of $f$ obtaining

\begin{eqnarray}
A&=&A_0+\sum_{j,k=1}^2 A_{j,k} f_{j,k}+\mathcal{O}(f_0),\nonumber\\
A_0&=&\kappa_2\lambda_2(\bar{c}_{0,1}^{+}c_{2,\Gamma_-}+c_{0,1}^-\bar{c}_{2,\Gamma_+}),\nonumber\\
A_{j,k}&=&\frac{1}{2}\kappa_2\lambda_2(c_{0,1}^{-}\bar{s}_{2,\Gamma_+}\Lambda_{j,k}^+-\bar{c}_{0,1}^+s_{2,\Gamma_-}\Lambda_{j,k}^-),\nonumber
\end{eqnarray}
\begin{eqnarray}
B&=&B_0+\sum_{j,k=1}^2 B_{j,k} f_{j,k}+\mathcal{O}(f_0),\nonumber\\
B_0&=&\kappa_2\lambda_1\bar{c}_{0,2}^+c_{2,\Gamma_-}+\kappa_1\lambda_2\bar{c}_{2,\Gamma_+}c_{0,2}^-,\nonumber\\
B_{j,k}&=&\frac{1}{2}(\kappa_1\lambda_2c_{0,2}^{-}\bar{s}_{2,\Gamma_+}\Lambda_{j,k}^+-\kappa_2\lambda_1\bar{c}_{0,2}^+s_{2,\Gamma_-}\Lambda_{j,k}^-),\nonumber
\end{eqnarray}

\begin{eqnarray}
C&=&C_0+\sum_{j,k=1}^2 C_{j,k} f_{j,k}+\mathcal{O}(f_0),\nonumber\\
C_0&=&\kappa_1\lambda_2\bar{c}_{0,1}^+c_{1,\Gamma_-}+\kappa_2\lambda_1\bar{c}_{1,\Gamma_+}c_{0,1}^-,\nonumber\\
C_{j,k}&=&\frac{1}{2}(\kappa_2\lambda_1c_{0,1}^{-}\bar{s}_{1,\Gamma_+}\Lambda_{j,k}^+-\kappa_1\lambda_2\bar{c}_{0,1}^+s_{1,\Gamma_-}\Lambda_{j,k}^-),\nonumber
\end{eqnarray}

\begin{eqnarray} 
D&=&D_0+\sum_{j,k=1}^2 D_{j,k} f_{j,k}+\mathcal{O}(f_0),\nonumber\\
D_0&=&\kappa_1\lambda_1(\bar{c}_{0,2}^{+}c_{1,\Gamma_-}+c_{0,2}^-\bar{c}_{1,\Gamma_+}),\nonumber\\
D_{j,k}&=&\frac{1}{2}\kappa_1\lambda_1(c_{0,2}^{-}\bar{s}_{1,\Gamma_+}\Lambda_{j,k}^+-\bar{c}_{0,2}^+s_{1,\Gamma_-}\Lambda_{j,k}^-),\nonumber
\end{eqnarray}

\begin{eqnarray}
Z&=&Z_0+\sum_{j,k=1}^2 Z_{j,k} f_{j,k}+\mathcal{O}(f_0),\nonumber\\
Z_0&=&\kappa_2\lambda_1\bar{c}_{1,\Gamma_+}c_{2,\Gamma_-}-\kappa_1\lambda_2c_{1,\Gamma_-}\bar{c}_{2,\Gamma_+},\nonumber\\
Z_{j,k}&=&\frac{1}{2}(\kappa_2\lambda_1 c_{2,\Gamma_-}\bar{s}_{1,\Gamma_+}-\kappa_1\lambda_2c_{1,\Gamma_-}\bar{s}_{2,\Gamma_+})\Lambda_{j,k}^+\nonumber\\
&-&\frac{1}{2}(\kappa_2\lambda_1 s_{2,\Gamma_-}\bar{c}_{1,\Gamma_+}-\kappa_1\lambda_2s_{1,\Gamma_-}\bar{c}_{2,\Gamma_+})\Lambda_{j,k}^-,\nonumber
\end{eqnarray}
where $\mathcal{O}(f_0)$ denotes terms proportional to  $f_0$.  It then follows
\begin{eqnarray}
\frac{X}{Z}&=&\frac{X_0}{Z_0}\left[1+\sum_{j,k=1}^2\left(\frac{X_{j,k}}{X_0}-\frac{Z_{j,k}}{Z_0}\right)f_{j,k}\right]+\mathcal{O}(f_0),\nonumber
\end{eqnarray}
where $X=\{A,B,C,D\}$.

 Substituting (\ref{matrix}), we obtain
\begin{eqnarray}
\mathcal{F}^{(1,3)}&=&\frac{A_0}{Z_0}\mathcal{F}^{(0,1)}-\frac{B_0}{Z_0}\mathcal{F}^{(0,2)}+\omega_1^{(1)}\mathcal{F}^{(0,1)}f_{2,1}+\omega_2^{(1)}\mathcal{F}^{(0,1)}f_{2,2},\label{F13}\\
\mathcal{F}^{(2,3)}&=&\frac{C_0}{Z_0}\mathcal{F}^{(0,1)}-\frac{D_0}{Z_0}\mathcal{F}^{(0,2)}+\omega_1^{(2)}\mathcal{F}^{(0,1)}f_{2,1}+\omega_2^{(2)}\mathcal{F}^{(0,1)}f_{2,2},\label{F23}
\end{eqnarray} 
where
\begin{eqnarray}
\omega_1^{(1)}&=&\frac{A_0}{Z_0}\left(\frac{A_{2,1}}{A_0}-\frac{Z_{2,1}}{Z_0}\right)+\frac{B_0}{Z_0}\left(\frac{B_{1,1}}{B_0}-\frac{Z_{1,1}}{Z_0}\right),\nonumber\\
\omega_2^{(1)}&=&\frac{A_0}{Z_0}\left(\frac{A_{2,2}}{A_0}-\frac{Z_{2,2}}{Z_0}\right)+\frac{B_0}{Z_0}\left(\frac{B_{1,2}}{B_0}-\frac{Z_{1,2}}{Z_0}\right),\nonumber\\
\omega_1^{(2)}&=&\frac{C_0}{Z_0}\left(\frac{C_{2,1}}{C_0}-\frac{Z_{2,1}}{Z_0}\right)+\frac{D_0}{Z_0}\left(\frac{D_{1,1}}{D_0}-\frac{Z_{1,1}}{Z_0}\right),\nonumber\\
\omega_2^{(2)}&=&\frac{C_0}{Z_0}\left(\frac{C_{2,2}}{C_0}-\frac{Z_{2,2}}{Z_0}\right)+\frac{D_0}{Z_0}\left(\frac{D_{1,2}}{D_0}-\frac{Z_{1,2}}{Z_0}\right).\nonumber
\end{eqnarray}
From eqns. (\ref{D+}), we get
\begin{eqnarray}
D_{+}(\phi_3^+-\phi_0^+)&=&\frac{8}{\kappa_1}\mathcal{F}^{(0,1)}c_{0,1}^-+\frac{8}{\kappa_2}\mathcal{F}^{(1,3)}c_{1,3}^-,\nonumber\\
D_{+}(\phi_1^+-\phi_2^+)&=&\frac{8}{\kappa_1}\mathcal{F}^{(0,1)}c_{0,1}^--\frac{8}{\kappa_2}\mathcal{F}^{(0,2)}c_{0,2}^-.\nonumber
\end{eqnarray}  
Introducing  solution 
(\ref{ansatz}) in the first eqn. above, we find 
\begin{eqnarray}
D_{+}(\phi_3^+-\phi_0^+)&=&D_+(\Gamma_++\Delta_+)=\nonumber\\
&=&\partial_x\Gamma_+D_{+}(\phi_1^+-\phi_2^+)+D_+\Delta_+=\nonumber\\
&=&\frac{8}{\kappa_1}\mathcal{F}^{(0,1)}c_{0,1}^-+\frac{8}{\kappa_2}\mathcal{F}^{(1,3)}c_{1,3}^-.\nonumber
\end{eqnarray}

Using  eqn. (\ref{F13}) in the above expression, and taking into account that  $\mathcal{F}^{(0,1)}$, $\mathcal{F}^{(0,2)}$, $\mathcal{F}^{(0,1)}f_{2,1}$ and  $\mathcal{F}^{(0,1)}f_{2,2}$ are independent, we arrive at the following conditions,
\begin{eqnarray}
&&\frac{c_{0,1}^-}{\kappa_1}(\partial_x\Gamma_+-1)-\frac{c_{1,\Gamma_-}}{\kappa_2}\frac{A_0}{Z_0}+\frac{g s_{0,2}^-}{4\kappa_2}\Lambda_{1,2}^++\frac{gs_{0,1}^-}{4\kappa_1}\Lambda_{1,1}^+=0,\nonumber\\
&&\frac{c_{0,2}^-}{\kappa_2}\partial_x\Gamma_+-\frac{c_{1,\Gamma_-}}{\kappa_2}\frac{B_0}{Z_0}-\frac{gs_{0,1}^-}{4\kappa_1}\Lambda_{2,1}^+-\frac{gs_{0,2}^-}{4\kappa_2}\Lambda_{2,2}^+=0,\label{Lambda+1}\\
&&\frac{c_{0,2}^-}{\kappa_2}\partial_x\Lambda_{1,1}^++\frac{c_{0,1}^-}{\kappa_1}\partial_x\Lambda_{2,1}^++\frac{gs_{0,2}^-}{4\kappa_2}\Lambda_{0}^+-\frac{c_{1,\Gamma_-}}{\kappa_2}\omega_1^{(1)}+\frac{s_{1,\Gamma_-}}{2\kappa_2}\left(\frac{A_0}{Z_0}\Lambda_{2,1}^-+\frac{B_0}{Z_0}\Lambda_{1,1}^-\right)=0,\nonumber\\
&&\frac{c_{0,2}^-}{\kappa_2}\partial_x\Lambda_{1,2}^++\frac{c_{0,1}^-}{\kappa_1}\partial_x\Lambda_{2,2}^+-\frac{gs_{0,1}^-}{4\kappa_1}\Lambda_{0}^+-\frac{c_{1,\Gamma_-}}{\kappa_2}\omega_2^{(1)}+\frac{s_{1,\Gamma_-}}{2\kappa_2}\left(\frac{A_0}{Z_0}\Lambda_{2,2}^-+\frac{B_0}{Z_0}\Lambda_{1,2}^-\right)=0.\nonumber
\end{eqnarray}
Moreover, the chirality condition on  (\ref{ansatz}) gives
\begin{eqnarray}
\bar{D}_-(\phi_3^+-\phi_0^+)=\bar{D}_-(\Gamma_++\Delta_+)=0,\nonumber
\end{eqnarray}
from where we obtain the following eqns.
\begin{eqnarray}
&&\frac{\bar{c}_{0,1}^+}{\lambda_1}\partial_y\Gamma_++\frac{g\bar{s}_{0,1}^+}{4\lambda_1}\Lambda_{1,1}^++\frac{g\bar{s}_{0,2}^+}{4\lambda_2}\Lambda_{1,2}^+=0,\nonumber\\
&&\frac{\bar{c}_{0,2}^+}{\lambda_2}\partial_y\Gamma_+-\frac{g\bar{s}_{0,1}^+}{4\lambda_1}\Lambda_{2,1}^+-\frac{g\bar{s}_{0,2}^+}{4\lambda_2}\Lambda_{2,2}^+=0,\nonumber\\
&&\frac{\bar{c}_{0,2}^+}{\lambda_2}\partial_y\Lambda_{1,1}^++\frac{\bar{c}_{0,1}^+}{\lambda_1}\partial_y\Lambda_{2,1}^++\frac{g\bar{s}_{0,2}^+}{4\lambda_2}\Lambda_{0}^+=0,\nonumber\\
&&\frac{\bar{c}_{0,2}^+}{\lambda_2}\partial_y\Lambda_{1,2}^++\frac{\bar{c}_{0,1}^+}{\lambda_1}\partial_y\Lambda_{2,2}^+-\frac{g\bar{s}_{0,1}^+}{4\lambda_1}\Lambda_{0}^+=0.\label{Lambda+2}
\end{eqnarray}
The two sets of eqns. namely,  (\ref{Lambda+1}) and  (\ref{Lambda+2}) give the following solutions,
\begin{eqnarray}
\Lambda_{1,1}^+&=&\Lambda_{2,2}^+=-\frac{8\mu_-}{g\eta_+\eta_-}\cos\left(\frac{x}{2}\right)\sin\left(\frac{y}{2}\right),\nonumber\\
\Lambda_{1,2}^+&=&\frac{8\mu_-}{g\eta_+\eta_-}\left(\frac{\lambda_2}{\lambda_1}\right)\sin\left(\frac{y}{2}\right),\nonumber\\
\Lambda_{2,1}^+&=&\frac{8\mu_-}{g\eta_+\eta_-}\left(\frac{\lambda_1}{\lambda_2}\right)\sin\left(\frac{y}{2}\right),\nonumber\\
\Lambda_{0}^+&=&-\frac{32\mu_-}{(g\eta_+\eta_-)^2}\sin\left(\frac{x}{2}\right)\left[\cos\left(\frac{y}{2}\right)(a+\cos x-\cos y)-2\mu_+\cos\left(\frac{x}{2}\right)\right],\nonumber
\end{eqnarray}
where
\begin{eqnarray}
\mu_{\pm}&=&\frac{\gamma_1}{\gamma_2}\pm\frac{\gamma_2}{\gamma_1}, \nonumber\\ a&=&\frac{1}{2}\left(\frac{\gamma_1^2}{\gamma_2^2}+\frac{\gamma_2^2}{\gamma_1^2}\right)+3, \nonumber\\ \eta_{\pm}&=&\mu_+-2\cos\left(\frac{x\pm y}{2}\right).\label{mu-a-eta}
\end{eqnarray} 
In order to determine the coefficients  $\Lambda^-$ we make use of 
\begin{eqnarray}
\bar{D}_{-}(\phi_3^--\phi_0^-)&=&\frac{8}{\lambda_1}\mathcal{F}^{(0,1)}\bar{c}_{0,1}^+-\frac{8}{\lambda_2}\mathcal{F}^{(1,3)}\bar{c}_{1,3}^+,\nonumber\\
\bar{D}_{-}(\phi_1^--\phi_2^-)&=&-\frac{8}{\lambda_1}\mathcal{F}^{(0,1)}\bar{c}_{0,1}^++\frac{8}{\lambda_2}\mathcal{F}^{(0,2)}\bar{c}_{0,2}^+,\nonumber
\end{eqnarray}
which are obtained from (\ref{xx}).  Introducing (\ref{ansatz})  in the first of these eqns. we find
\begin{eqnarray}
\bar{D}_{-}(\phi_3^--\phi_0^-)&=&\bar{D}_-(\Gamma_-+\Delta_-)=\nonumber\\
&=&\partial_y\Gamma_-\bar{D}_{-}(\phi_1^--\phi_2^-)+\bar{D}_-\Delta_-=\nonumber\\
&=&\frac{8}{\lambda_1}\mathcal{F}^{(0,1)}\bar{c}_{0,1}^+-\frac{8}{\lambda_2}\mathcal{F}^{(1,3)}\bar{c}_{1,3}^+.\nonumber
\end{eqnarray} 

Using  eqn. (\ref{F13}) in the above expression and  taking into account that  $\mathcal{F}^{(0,1)}$, $\mathcal{F}^{(0,2)}$, $\mathcal{F}^{(0,1)}f_{2,1}$ and $\mathcal{F}^{(0,1)}f_{2,2}$ are independent, we arrive  at the following expressions,
\begin{eqnarray}
&&\frac{\bar{c}_{0,1}^+}{\lambda_1}(\partial_y\Gamma_-+1)-\frac{\bar{c}_{1,\Gamma_+}}{\lambda_2}\frac{A_0}{Z_0}+\frac{g \bar{s}_{0,1}^+}{4\lambda_1}\Lambda_{1,1}^-+\frac{g\bar{s}_{0,2}^+}{4\lambda_2}\Lambda_{1,2}^-=0,\nonumber\\
&&\frac{\bar{c}_{0,2}^+}{\lambda_2}\partial_y\Gamma_--\frac{\bar{c}_{1,\Gamma_+}}{\lambda_2}\frac{B_0}{Z_0}-\frac{g\bar{s}_{0,1}^+}{4\lambda_1}\Lambda_{2,1}^--\frac{g\bar{s}_{0,2}^+}{4\lambda_2}\Lambda_{2,2}^-=0,\label{Lambda-1}\\
&&\frac{\bar{c}_{0,2}^+}{\lambda_2}\partial_y\Lambda_{1,1}^-+\frac{\bar{c}_{0,1}^+}{\lambda_1}\partial_y\Lambda_{2,1}^-+\frac{g\bar{s}_{0,2}^+}{4\lambda_2}\Lambda_{0}^--\frac{\bar{c}_{1,\Gamma_+}}{\lambda_2}\omega_1^{(1)}-\frac{\bar{s}_{1,\Gamma_+}}{2\lambda_2}\left(\frac{A_0}{Z_0}\Lambda_{2,1}^++\frac{B_0}{Z_0}\Lambda_{1,1}^+\right)=0,\nonumber\\
&&\frac{\bar{c}_{0,2}^+}{\lambda_2}\partial_y\Lambda_{1,2}^-+\frac{\bar{c}_{0,1}^+}{\lambda_1}\partial_y\Lambda_{2,2}^--\frac{g\bar{s}_{0,1}^+}{4\lambda_1}\Lambda_{0}^--\frac{\bar{c}_{1,\Gamma_+}}{\lambda_2}\omega_2^{(1)}-\frac{\bar{s}_{1,\Gamma_+}}{2\lambda_2}\left(\frac{A_0}{Z_0}\Lambda_{2,2}^++\frac{B_0}{Z_0}\Lambda_{1,2}^+\right)=0.\nonumber
\end{eqnarray}

The chiral condition
\begin{eqnarray}
D_+(\phi_3^--\phi_0^-)=D_+(\Gamma_-+\Delta_-)=0,\nonumber
\end{eqnarray}
leads us to
\begin{eqnarray}
&&\frac{c_{0,1}^-}{\kappa_1}\partial_x\Gamma_-+\frac{gs_{0,1}^-}{4\kappa_1}\Lambda_{1,1}^-+\frac{gs_{0,2}^-}{4\kappa_2}\Lambda_{1,2}^-=0,\nonumber\\
&&\frac{c_{0,2}^-}{\kappa_2}\partial_x\Gamma_--\frac{gs_{0,1}^-}{4\kappa_1}\Lambda_{2,1}^--\frac{gs_{0,2}^-}{4\kappa_2}\Lambda_{2,2}^-=0,\nonumber\\
&&\frac{c_{0,2}^-}{\kappa_2}\partial_x\Lambda_{1,1}^-+\frac{c_{0,1}^-}{\kappa_1}\partial_x\Lambda_{2,1}^-+\frac{gs_{0,2}^-}{4\kappa_2}\Lambda_{0}^-=0,\nonumber\\
&&\frac{c_{0,2}^-}{\kappa_2}\partial_x\Lambda_{1,2}^-+\frac{c_{0,1}^-}{\kappa_1}\partial_x\Lambda_{2,2}^--\frac{gs_{0,1}^-}{4\kappa_1}\Lambda_{0}^-=0.\label{Lambda-2}
\end{eqnarray}

Solving (\ref{Lambda-1}) and  (\ref{Lambda-2}) for  $\Lambda^-$, we find
\begin{eqnarray}
\Lambda_{1,1}^-&=&\Lambda_{2,2}^-=\frac{8\mu_-}{g\eta_+\eta_-}\cos\left(\frac{y}{2}\right)\sin\left(\frac{x}{2}\right),\nonumber\\
\Lambda_{1,2}^-&=&-\frac{8\mu_-}{g\eta_+\eta_-}\left(\frac{\kappa_2}{\kappa_1}\right)\sin\left(\frac{x}{2}\right),\nonumber\\
\Lambda_{2,1}^-&=&-\frac{8\mu_-}{g\eta_+\eta_-}\left(\frac{\kappa_1}{\kappa_2}\right)\sin\left(\frac{x}{2}\right),\nonumber\\
\Lambda_{0}^-&=&-\frac{32\mu_-}{(g\eta_+\eta_-)^2}\sin\left(\frac{y}{2}\right)\left[\cos\left(\frac{x}{2}\right)(a-\cos x+\cos y)-2\mu_+\cos\left(\frac{y}{2}\right)\right],\nonumber
\end{eqnarray}
where $\mu_{\pm}$, $a$ and  $\eta_{\pm}$ are given in  (\ref{mu-a-eta}).

\end{document}